\documentclass[12pt]{iopart}

\usepackage{amsgen}
\usepackage{amssymb}
\usepackage{setstack}
\usepackage{bbold}
\usepackage{mathrsfs}
\usepackage{graphicx}
\usepackage{multicol}
\usepackage{color}

\def\be{\begin{equation}}
\def\ee{\end{equation}}
\def\bea{\begin{eqnarray}}
\def\eea{\end{eqnarray}}

\begin{document}

\title{Persistent Black Holes in Bouncing Cosmologies}

\author{Timothy Clifton$^1$, Bernard Carr$^1$ and Alan Coley$^{2}$}
\address{$^1$ School of Physics and Astronomy, Queen Mary University of London, UK.\\
$^2$ Department of Mathematics and Statistics, Dalhousie University, Canada.}

\eads{\mailto{t.clifton@qmul.ac.uk, b.j.carr@qmul.ac.uk, aac@mathstat.dal.ca}}


\begin{abstract}

In this paper we explore the idea that black holes can persist in a universe that collapses to a big crunch and then bounces into a new phase of expansion. We use a scalar field to model the matter content of such a universe {near the time} of the bounce, and look for solutions that represent a network of black holes within a dynamical cosmology. We find exact solutions to Einstein's constraint equations that provide the geometry of space at the minimum of expansion and that can be used as initial data for the evolution of hyperspherical cosmologies. These solutions illustrate that there exist models in which multiple distinct black holes can persist through a bounce, and allow for concrete computations of quantities such as the black hole filling factor. We then consider solutions in flat cosmologies, as well as in higher-dimensional spaces (with up to nine spatial dimensions). We derive conditions for the black holes to remain distinct {(i.e. avoid merging) and hence} persist into the new expansion phase. Some potentially interesting consequences of these models are also discussed.

\end{abstract}

\section{Introduction}

Bouncing cosmologies, in which the Universe collapses to a minimum radius before returning to expansion, have a long history \cite{lemaitre}. 
As stressed in a previous paper \cite{cc},  it is interesting to consider whether black holes can persist through a cosmological bounce and what effect they would have on the large-scale dynamics and on each other.
One {can} divide {such} black holes into two classes: (i) those that persist in a universe that recollapses to a big crunch and then bounces into a new expansion phase; and (ii) those that are generated by the high density of matter at the bounce. {Following Ref.~\cite{cc}, these objects will be} referred to as ``pre-crunch black holes'' (PCBHs) and ``big-crunch black holes'' (BCBHs), respectively. It is interesting to consider whether there would be any observational way of distinguishing {them} from the primordial black holes  {(PBHs)} that formed {shortly after the bounce, or during the early stages of the standard hot big bang model without a bounce}. It is also  {important to know whether black holes  really can} persist through a bounce in the way that is being envisaged.

The precise properties of a cosmic bounce depend upon the way in which it is generated. {Many mechanisms} have been proposed for this. These include classical {effects associated with} a cosmological constant \cite{lemaitre}, a scalar field \cite{star}, higher-derivative effects \cite{novello} or other modified theories of gravity \cite{tors1,tors2}. {There are also} semi-quantum gravitational effects associated with string theory \cite{string}, loop quantum gravity \cite{sing,sing2,sing3}, quantum gravity condensates \cite{oriti}, and the ``pre-big-bang'' scenario \cite{BHScorr}. There have also been some claims that scale-invariant fluctuations generated during a cosmological collapse phase could turn into scale-invariant curvature fluctuations after a bounce \cite{cai0}, although in a fully quantum gravitational bounce there may be conceptual {problems with} these approaches \cite{ashtekar}. For a recent review of bouncing cosmologies, see Ref.~\cite{brand}. The scalar-field-driven bounces we consider here are probably the simplest {models that can accommodate} a bounce, and allow us to consider the problem of persistent black holes in an exact and fully relativistic way.

{Besides} the mechanism of the bounce itself, there is also the tricky mathematical issue of what is meant by a black hole in a Universe that is not eternally expanding.  This issue is further complicated by the lack of any asymptotic spatial infinity in most cosmological models. For a {PBH} that forms after the big bang, it is usually understood that the black hole singularity is part of any future big crunch singularity (if one should occur). {However,} this cannot apply to PCBHs or BCBHs if the cosmological bounce is non-singular.  Without knowledge of the 
global structure of the spacetime, it is impossible to locate the event horizon, which is defined as the boundary of the past of future null infinity. A more satisfactory approach is therefore to use the concept of the  {black hole} apparent horizon, {which is} defined as the outermost Marginally Outer Trapped Surface (MOTS) \cite{MOTS,hayward}. This approach is now used routinely in numerical work and we will also adopt it here. It has even been argued that Hawking radiation is associated with the apparent horizon \cite{cai}. For a more detailed discussion of these issues, see Ref.~\cite{faraoni}. 

A matter of particular interest for models of this type concerns what happens to the number density of black holes as the minimum of contraction is approached. One could speculate about the maximum {fraction} of space that can be filled by PCBHs or {about} the number of BCBHs that form during the collapsing stage. These quantities would depend on the density of the matter at the bounce and there are then various interesting constraints that can be imposed on the fraction of the Universe's mass density that resides in black holes as a function of mass \cite{cc}. If the black holes are randomly distributed, one would expect a process of hierarchical merging to occur, in which progressively larger apparent horizons form around groups of holes, so that the characteristic hole size steadily increases. However, if the holes were distributed uniformly (eg. on a lattice), then pre-crunch hierarchical merging would be eliminated and they might maintain their individuality. In this case, one could investigate the maximum number of black holes that could fit through the bounce.

In this paper, we {study} this scenario {in detail} by considering models that contain a regular lattice of black holes and a scalar field at the maximum compression. These restrictions are primarily motivated by mathematical simplicity, as we find that they {yield} a number of exact solutions to the Einstein constraint equations. The {associated} initial data {then prescribe} a unique evolution in both the expansion and collapse {directions}. They therefore constitute a {proper} cosmological model. Although in general the evolution will need to be determined numerically, the initial data itself can be used to analyse a number of the problems outlined above. In particular, the positions of the apparent horizons can be determined and used to calculate the distance between neighbouring black holes. This information can be used to determine the fraction of space that is filled by black holes as a function of the maximum energy density at the bounce. It also unambiguously demonstrates that black holes can persist through a cosmological bounce.

Another {interesting} consideration for such a scenario is whether the matter content {of} the expanding phase is identical to that of the collapsing phase. In general, one might expect entropy to increase from one cycle to the next \cite{tolman,penrose}. This implies the production of radiation or the creation of black holes, which can be thought of as the objects with the highest entropy allowed in nature \cite{bousso}. We address this issue in more detail in a separate paper \cite{future}. 

The plan of this paper is as follows. In Sec.~2 we derive the form of the constraint equations in a time-symmetric scalar field-dominated model, focussing particularly on the geometry of the spatial hypersurface at the bounce. In Secs.~3 and 4 we consider black holes in hyperspherical and spatially flat cosmological models, respectively. In Sec.~5 we consider black holes in higher-dimensional bouncing cosmologies, a possibility also emphasized in Ref.~\cite{cc}. In Sec.~6 we draw some general conclusions. Appendix~A {discusses} solutions for the initial data in which the scalar field is stationary at the maximum compression, {these relating} to multiple Janis-Newman-Winicour-like black hole solutions. {Appendix~B discusses} solutions with a non-vanishing scalar potential.

\section{Constraint and evolution equations} 
\label{sec:constraints}

In this {section} we will present the Einstein field equations (EFEs) in the presence of radiation and a scalar field, as relevant for studying the geometry of space-time in the vicinity of a bounce. This follows from the standard approach of using a scalar field in early-universe Friedmann-Lema\^{i}tre-Robertson-Walker (FLRW) cosmology in order to model the effects of new physics. In our case, however, we will not assume anything about the existence of global symmetries (approximate or otherwise). Instead, we will give the general form for the constraint and evolution equations, and describe how a bouncing cosmology with black holes can be formulated as an initial value problem, with initial data prescribed at the bounce.

Just as in standard FLRW cosmology, the scalar field may be expected to dominate at early times but become negligible compared to the energy and momentum in radiation and dust at late times. This means that if one considers a regular array of black holes and a scalar field, the late-time evolution of the space might approach that of a space just containing an array of black holes (i.e. the space should forget that it once contained a dominant scalar field). This is mathematically very interesting, because for a scalar field with a negative coupling constant there exists a time-symmetric initial value problem for an array of black holes which expand away from each other as the space evolves from the time-symmetric surface. The existence of time symmetry means that the constraint equations on the initial surface can be manipulated into a linear form, allowing for the initial data for arbitrary many black holes to be constructed by simple superposition.

\subsection{Field equations and time-symmetric initial data}

Let us consider the EFEs with a scalar field:
\be
\label{fe}
G_{ab} = \mu T^{\varphi}_{ab} \,  + \, T^{\gamma}_{ab} \, ,
\ee
where $\mu$ is a coupling constant (necessarily negative for a bounce) and {we have chosen units with} $8 \pi G = c=1$. The quantity $T^{\varphi}_{ab}$ is the energy-momentum tensor of the scalar field, and is given by
\be
T^{\varphi}_{ab} = \nabla_a \varphi \nabla_b \varphi -\frac{1}{2} g_{ab}   \nabla_c \varphi \nabla^c \varphi - g_{ab}  V(\varphi) \, ,
\ee
where $V(\varphi)$ is the scalar potential. The quantity $T^{\gamma}_{ab}$ is the energy-momentum tensor of an additional radiation fluid, which is assumed to be separately conserved. The contracted second Bianchi identity then gives the scalar propagation equation as
\be
\label{boxvarphi}
\square \varphi = \frac{dV}{d\varphi} \, ,
\ee
where $\square \equiv \nabla^a \nabla_a$ is the covariant d'Alembertian operator.

If we now choose a time-like vector field, $u^a$, we can decompose $T^{\varphi}_{ab}$ into energy density, isotropic pressure and momentum density:
\bea
\label{rho}
\rho^{\varphi} &\equiv& u^a u^b T^{\varphi}_{ab} = \frac{1}{2} \dot{\varphi}^2 + \frac{1}{2} D^a \varphi D_a \varphi  +V(\varphi)\, \\
\label{p} p^{\varphi} &\equiv& \frac{1}{3} h^{ab} T^{\varphi}_{ab} = \frac{1}{2} \dot{\varphi}^2 - \frac{1}{6} D^a \varphi D_a \varphi -V(\varphi) \,\\  
\label{j}
j_a^{\varphi} &\equiv& - h_a^{\phantom{a} b} u^c T^{\varphi}_{bc} = - \dot{\varphi} \, D_a \varphi \, .
\eea
Here an over-dot denotes $u^a \nabla_a$, and we have introduced $h_{ab} \equiv g_{ab} +u_a u_b$ and $D_a \equiv h_a^{\phantom{a} b} \nabla_b$ as the projection operator and the projected covariant derivative on the set of spaces orthogonal to $u^a$. Likewise, the energy density and isotropic pressure of a perfect fluid of radiation are given by
\bea
\rho^{\gamma} &\equiv& u^a u^b T^{\gamma}_{ab} \, , \qquad p^{\gamma} \equiv \frac{1}{3} h^{ab} T^{\gamma}_{ab} \, ,
\eea
where $p^{\gamma} = \frac{1}{3} \rho^{\gamma}$ and the momentum density is $j_a^{\gamma} \equiv - h_a^{\phantom{a} b} u^c T^{\gamma}_{bc} = 0$ (by virtue of the fluid being radiation and perfect, respectively).

Using the appropriate embedding equations, together with the EFEs (\ref{fe}), we can write the Hamiltonian and momentum constraint equations for the space-time as
\bea
\label{hc}
&&\mathcal{R} + K^2 - K_{ab} K^{ab} = 2 u^a u^b G_{ab} = 2 \mu \rho^{\varphi} + 2 \rho^{\gamma}  \,  \\
\label{mc}
&&D_b K^b_{\phantom{b} a} - D_a K = - h^b_{\phantom{b} a} u^c G_{bc} = \mu j_a^{\varphi} \, ,
\eea
where $K_{ab} \equiv - h_a^{\phantom{a} c} h_b^{\phantom{b} d} \nabla_{(c} u_{d)}$ is the extrinsic curvature of the hypersurface we are considering, $K \equiv K^a_{\phantom{a} a}$ and $\mathcal{R}$ denotes the Ricci scalar of this hypersurface (assuming $u^a$ is irrotational). These equations must be satisfied on the initial hypersurface, in order to have a solution to the EFEs. The corresponding evolution equations for the geometry will be presented in Sec. \ref{sec:time} below.

It now remains to decompose the conservation equations for the scalar field and radiation fluids. For the radiation field this yields
\bea
\label{radcon}
D_a \rho^{\gamma} + 4 \rho^{\gamma} \dot{u}_a =0 \, , \qquad
\dot{\rho}^{\gamma} = \frac{4}{3} K \rho^{\gamma}
\eea
as the constraint and evolution equations for $\rho^{\gamma}$, respectively. Here we have chose a  $u^a$ that is comoving with the radiation fluid. For the scalar field, the propagation equation (\ref{boxvarphi}) can be decomposed into evolution and constraint equations with respect to the vector field $u^a$. To do this, we first define
\be
\label{varphivar}
 \psi_a \equiv D_a \varphi \, .
\ee
We can then decompose Eq. (\ref{boxvarphi}) into the following first-order evolution equations:
\bea
\label{evo1}
\dot{\Pi} &=& D_a \psi^a+ K \Pi + \dot{u}_a \psi^a + V^{\prime}(\varphi) \, \\
\dot{\psi}_a &=& D_a \Pi + \dot{u}_a \Pi + u_a \dot{u}^b \psi_b \, \\
\dot{\varphi} &=& \Pi \, ,
\label{evo3}
\eea
where the last of these {defines the quantity $\Pi$}. Eq.~(\ref{varphivar}) is then the only constraint equation, while Eqs.~(\ref{evo1})-(\ref{evo3}) constitute the evolution equations for $\Pi$, $\psi_a$ and $\varphi$. This completes our formulation of the initial value problem in terms of constraint and evolution equations, as required for this study.

Let us now consider time-symmetric initial data. In this case the initial data must {satisfy} $K_{ab}=0$. From Eqs.~(\ref{j}) and (\ref{mc}), this immediately implies
\be
\dot{\varphi} D_a \varphi =0 \, ,
\ee
so that either $\dot{\varphi}=0$ or $D_a \varphi=0$ or both on the initial hyper-surface. If both are true, then the scalar field evolution equations show that $\varphi =0$ at all points in space and time, which is the trivial case in which there is no scalar field. The case where $\dot{\varphi}=0$ is considered in the Appendices. If $V=0$, then initial data of this type can be shown to include cosmological solutions with multiple Janis-Newman-Winicour-like black holes (see Appendix A). If $V(\varphi)$ is an exponential function, then exact solutions can also be found (see Appendix B). However, the remaining case, where $D_a \varphi=0$, is a more interesting one for cosmology and will form the basis of the study that follows.

If $D_a \varphi =0$, then $\varphi$ is constant on the initial hypersurface and we can infer from Eqs.~(\ref{rho}) and (\ref{p}) that $\rho^{\varphi} = \frac{1}{2} \Pi^2 +V$ and $p^{\varphi} = \frac{1}{2} \Pi^2-V$. Time-symmetry on the initial hypersurface then requires $\dot{\rho}^{\varphi}= \Pi (\dot{\Pi} +V^{\prime})=0$ and $\dot{p}^{\varphi} = \Pi (\dot{\Pi} -V^{\prime})=0$. As we require $\Pi \neq 0$ in order to have a non-vacuum solution, the time-symmetry condition implies $\dot{\Pi}=V^{\prime}=0$, which is compatible with the scalar field evolution equation (\ref{evo1}). In this case, we therefore have a solution where $\varphi$ is initially homogeneous but $\dot{\varphi}$ is inhomogeneous and $\ddot{\varphi}$ vanishes. We are also forced to {require} $V$ to be a constant, {so that it acts} like a cosmological constant. We therefore neglect $V$ in what follows, as the cosmological constant is not expected to play an important role
{near the time of the} bounce.

The remaining equation that needs to be satisfied in order to have a solution to the Einstein-scalar system is the Hamiltonian constraint (\ref{hc}). When $D_a \varphi = V =0$, this can be written as
\be
\label{constraintR}
\mathcal{R} = \mu \Pi^2 + 2 \rho^{\gamma} \, .
\ee
The quantity $\Pi$ is not involved in the scalar field constraint equation and can therefore be specified freely. Once we have a solution for the geometry of an initial hypersurface that satisfies Eq.~(\ref{constraintR}) for a given $\Pi$,  we have a complete solution to all the constraint equations and hence the full initial data required to determine the entire evolution.

To find explicit solutions with constant conformal curvature, let us make the following ansatz for the line-element of the time-symmetric initial hypersurface:
\be
\label{le}
ds^2 = \Omega^4 d\bar{s}^2 \, ,
\ee
where $d\bar{s}^2$ is the line-element for a 3-space of constant  curvature. Standard results from conformal geometry then give
\be
\label{conformal}
\mathcal{R} = \frac{\bar{\mathcal{R}}}{\Omega^4} - \frac{8}{\Omega^5} \bar{D}^2 \Omega \, ,
\ee
where $\bar{\mathcal{R}}=\,$constant is the Ricci scalar of the conformal space and $\bar{D}$ is the covariant derivative on the conformal space. Combining Eqs.~(\ref{constraintR}) and (\ref{conformal}) then gives 
\be
\label{constraint}
\bar{D}^2 \Omega - \frac{\bar{\mathcal{R}}}{8} \Omega=  -\frac{(\mu \Pi^2 + 2 \rho^{\gamma})}{8}   \Omega^5  \, .
\ee
If $\mu\Pi^2 + 2 \rho^{\gamma}=0$, so that we are considering a vacuum space-time, then this equation is linear in $\Omega$, which is promising if we are interested in looking for a multi-black hole solution \cite{clifton1}-\cite{num0}. Although Eq.~(\ref{constraint}) is not linear in $\varphi$, it can be made so by a suitable choice of $\Pi$. It is this equation that we will use to solve for black holes in a hyperspherical cosmology in Section \ref{sec:sphere}, in a spatially flat cosmology in Section \ref{sec:flat} and in a higher-dimensional cosmology in Section \ref{sec:higher}. Given the EFEs, it can be seen that we can only obtain a bounce when the matter source violates the null energy conditions (e.g. when $\mu <0$ and $V=0$, which describes a massless scalar field with a negative energy density). However, bounces can be more naturally obtained with normal matter in modified gravity theories \cite{brand}.

\subsection{Time evolution of instantaneously-static models}
\label{sec:time}

In subsequent sections we will present exact solutions of the constraint equation (\ref{constraint}) that can be used to develop an exact 4-dimensional dynamical solutions of the EFEs. These solutions, while highly idealised, describe realisations of the scenario in which multiple distinct black holes persist through a cosmological bounce. Therefore, they both demonstrate the existence of solutions that could be generalised to more realistic models {and} provide space-time geometries in which specific computations can be made. However, in order to study the full time-evolution of these models we need to solve the full EFEs and not just the conservation and constraint equations. 

Let us choose coordinates so that the full 4-dimensional space-time is given by the following line-element:
\be
\label{dynmetric}
ds^2 = -A^2(t,x^{\gamma}) dt^2 +  h_{\alpha \beta} (t,x^{\gamma}) dx^{\alpha} dx^{\beta}.
\ee
An evolving solution would be a metric of this form solving both the constraint and the evolution equations. For a dynamical model with a bounce at the time-symmetic surface at $t=t_0$, the 3-metric we are interested in is
\be
\label{hypmetric}
h_{\alpha \beta} (t_0,x^{\gamma}) dx^{\alpha} dx^{\beta} = \Omega^4(x^{\gamma})d\bar{s}^2 \, .
\ee
We can then obtain an exact {\em{space-time}} by solving the evolution equations
\bea
\mathcal{L}_t h_{\alpha \beta} &=& -2 A K_{\alpha \beta}
\eea
and
\bea
\mathcal{L}_t K_{\alpha \beta} &=& - D_{\alpha} D_{\beta} A + A \left( \mathcal{R}_{\alpha \beta} - 2 K_{\alpha \gamma} K^{\gamma}_{\phantom{\gamma} \beta} + K \, K_{\alpha \beta} \right) \nonumber \\ &&+ \frac{A}{2} \left( p^{\varphi} - \rho^{\varphi} - 2 \Pi^{\varphi}_{\alpha \beta} - \frac{2}{3}\rho^{\gamma} \right) h_{\alpha \beta} \, ,
\eea
where $t^{a} \equiv A u^{a}$ {and} $\Pi^{\varphi}_{a b} = (h_a^{\phantom{a} c} h_b^{\phantom{b} d} - \frac{1}{3} h_{ab} h^{cd}) T^{\varphi}_{cd}$ is the anisotropic stress of the scalar field. As the initial value problem is well-posed, once appropriate initial values for the metric and extrinsic curvature functions have been specified, we are guaranteed a unique solution. In general, this will need to be obtained numerically \cite{num0}-\cite{num5}, although approximate solutions may also be possible \cite{num5}-\cite{curlB}.

At this point, to fix ideas, it may be useful to compare the inhomogeneous models we are constructing to analogous FLRW models. For an isotropic and spatially homogeneous geometry, {we} have $A=1$ and $h_{\alpha \beta} dx^{\alpha} dx^{\beta} = a^2(t) d \bar{s} ^2$, where $a(t)$ is the scale factor that describes the cosmic evolution in its entirety. In the massless scalar field FLRW models with dust, radiation, and spatial curvature there can of course exist a bounce, and the 
corresponding solution for $a(t)$ can be easily obtained. In the absence of energy exchange between these fluids, the bounce at $t=t_0$ is generically expected to be time-symmetric. This means that at the minimum of the bounce, the geometry described by the conformal factor $\Omega^2 = a(t_0)$ is instantaneously static. The time-symmetric initial data we considered above is the generalisation of this situation to inhomogeneous space-times that are capable of harbouring black holes. Just as the FLRW solution moves away from the minimum of expansion into a regular phase of expansion, we expect our initial data to evolve into a regular universe in which the scalar field dilutes, and the radiation and dust fields eventually come to dominate. That is, a set of instantaneously static initial data should not be expected to lead to a static space-time, but rather a dynamical cosmological model. This will be explained further in the sections that follow.

\section{Black holes in a hyperspherical cosmology}
\label{sec:sphere}

Let us start by considering hyperspherical cosmological models. We study this first not because it is more likely to be associated with a bounce, but because it is mathematically simpler when formulated in terms of an initial value problem. For a hyperspherical cosmology, we take the conformal line-element in Eq. (\ref{le}) to be
\be
\label{hs}
d\bar{s}^2 = dr^2 + \sin^2 r \left( d\theta^2 + \sin^2 \theta \, d \phi^2 \right) \, .
\ee
The scalar curvature of this space is $\bar{\mathcal{R}}=6$, so the Hamiltonian constraint (\ref{constraint}) becomes
\be
\bar{D}^2 \Omega = \left( \frac{3}{4} -\frac{1}{8} (\mu \Pi^2 +2 \rho^{\gamma}) \Omega^4 \right) \Omega   \, .
\label{ham}
\ee
If we now choose the initial value of $\Pi$ such that
\be
\label{c}
\Pi^2 = \frac{8}{\mu \Omega^4} \left( \frac{3}{4} - \kappa \right) -\frac{2 \rho^{\gamma}}{\mu} \, ,
\ee
where $\kappa$ is a constant, then the constraint equation becomes
\be
\label{helmholtz}
\bar{D}^2 \Omega = \kappa \, \Omega \, ,
\ee
so that $\kappa > 3/4$ if both $\mu <0$ and $\rho^{\gamma}=0$. This is the Helmholtz equation on a 3-sphere, which has well-known solutions. It is linear in $\Omega$, so we can construct multi-black hole solutions by simple superposition. If $\kappa = 3/4$ it can be seen that the energy densities in the scalar field and radiation must sum to zero on the initial hypersurface.

\subsection{Solutions for the conformal factor}

There exist solutions to Eq. (\ref{helmholtz}) of the following form:
\be
\label{omi}
\Omega_i (r) = \alpha_i \, \frac{\cos (\sqrt{1- \kappa} \, r)}{\sin r} + \gamma_i \, \frac{\sin (\sqrt{1- \kappa} \, r)}{\sin r} \, ,
\ee
where $\alpha_i$ and $\gamma_i$ are constants. These solutions diverge at $r=0$, which we take to be the location of the source. In order for the geometry of the space to be smooth at all other points, we require the first derivative of $\Omega$ to be single-valued. Applying this condition at $r=\pi$ implies $\gamma_i = - \alpha_i \cot ( \sqrt{1- \kappa}\,  \pi)$, so Eq. (\ref{omi}) gives
\be
\label{cont0}
\Omega_i(r) = \alpha_i \frac{\sin (\sqrt{1-\kappa}  \, (\pi-r))}{\sin (\sqrt{1-\kappa}  \, \pi) \sin r} \, .
\ee
This equation gives the contribution to the conformal factor for a single point-like source at $r=0$. For $N$ sources, located at arbitrary positions on the hypersphere, we can write the corresponding solution as
\be
\label{cont}
\Omega (r, \theta, \phi) = \sum_{i=1}^N \alpha_i \frac{\sin (\sqrt{1-\kappa}  \, (\pi-r_i))}{\sin (\sqrt{1-\kappa}  \, \pi) \sin r_i} \, ,
\ee
where $r_i$ should be understood as the radial coordinate in Eq.~(\ref{hs}), after rotation of the hypersphere so that the $i$th source is located at $r_i=0$.

\newpage
\noindent
These solutions, corresponding to $N$ point sources, are characterized by:
\begin{itemize}
\item Complete freedom in the positions of the sources on a reference 3-sphere.
\item One number to be specified for each black hole ($\alpha_i$).
\item A single number ($\kappa$) that sets the magnitude of the scalar field energy density.
\item A single number ($\mu$) for the coupling strength of the scalar field.
\end{itemize}
It now remains to determine the relationship between the constants $\alpha_i$ and the proper mass of each of the sources.

\subsection{Proper mass of sources}
\label{spheremass}

To determine the proper mass of one of the sources, we start by taking the $r_i \rightarrow 0$ limit of Eq.~(\ref{cont}). This gives
\be
ds^2 \rightarrow \left( \frac{\alpha_i}{r_i} +A_i \right)^4 \left( dr_i^2 +r_i^2 (d\theta^2 + \sin^2 \theta \, d \phi^2 ) \right) \, ,
\ee
where
\be
A_i \equiv - \frac{\alpha_i \sqrt{1-\kappa}}{ \tan (\sqrt{1-\kappa}  \, \pi)} + \sum_{j\neq i} \alpha_j \frac{\sin (\sqrt{1-\kappa}  \, (\pi-r_{ij}))}{\sin (\sqrt{1-\kappa}  \, \pi) \sin r_{ij}} \, 
\ee
and $r_{ij}$ is the coordinate distance to the $j$th source from the one located at $r_i=0$. We then define a new radial coordinate by $r^{\prime}_i \equiv \alpha_i^2 / r_i$, such that
\be
ds^2 \rightarrow \left(  1 + \frac{4\alpha_i  A_i}{r_i^{\prime}} \right) \left( dr^{\prime 2}_i +r_i^{\prime 2} (d\theta^2 + \sin^2 \theta \, d \phi^2 ) \right) \, .
\ee
This solution is dual to the original one, in the sense that $r=0$ and infinity have been interchanged; this is because each black hole is represented in the initial data by an Einstein-Rosen-type bridge such that the dual solution is just the reflection of the original one in the apparent horizon. This new geometry can be compared with the Schwarzschild metric in the limit $r \rightarrow \infty$,
\be
\label{schwarz}
ds^2 \rightarrow \left( 1 + \frac{2 m}{r} \right) \left( dr^2 + r^2 (d\theta^2 + \sin^2 \theta d \phi^2) \right) \, ,
\ee
which gives the proper mass of the $i$th source as
\be
m_i = 2 \alpha_i A_i = - \frac{2 \alpha_i^2 \sqrt{1-\kappa}}{ \tan (\sqrt{1-\kappa} \pi)} + 2 \sum_{j\neq i} \alpha_i \alpha_j \frac{\sin (\sqrt{1-\kappa}  \, (\pi-r_{ij}))}{\sin (\sqrt{1-\kappa}  \, \pi) \sin r_{ij}} \, .
\ee
Note also that the mass of each source is a function of every $\alpha_j$ and the position of every other source. This mass can be either positive or negative, depending on the value of constants $\kappa$, $\alpha_i$ and $r_{ij}$. Note that the expression above is regular at $\kappa \geqslant 1$, as in these cases the trigonometric functions merely turn into hyperbolic ones.

\subsection{Example configuration: Eight regularly-spaced masses}
\label{sec:8mass}

\begin{figure}[tbh!]
\begin{center}
\includegraphics[width=10cm]{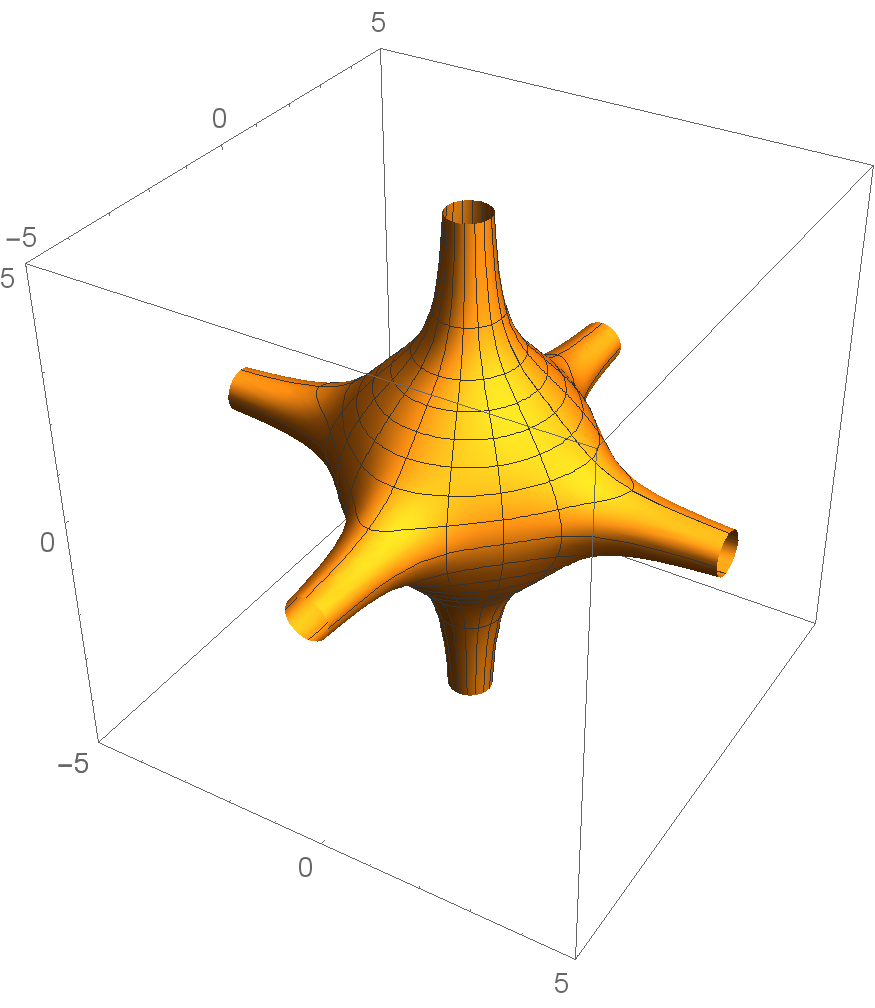}
\caption{A graphical representation of the scale factor $\Omega$ in $(x,y,z)$ space for the eight-mass configuration. The value of $\Omega$ in this plot is proportional to the distance of the surface from the origin, and it is the 2-dimensional surface at $r=\pi/2$ that is displayed. We have taken $m=1$ for each black hole and $\kappa=0.54$.}
\centering{}\label{om8}
\end{center}
\end{figure}

\begin{figure}[tbh!]
\begin{center}
\includegraphics[width=10cm]{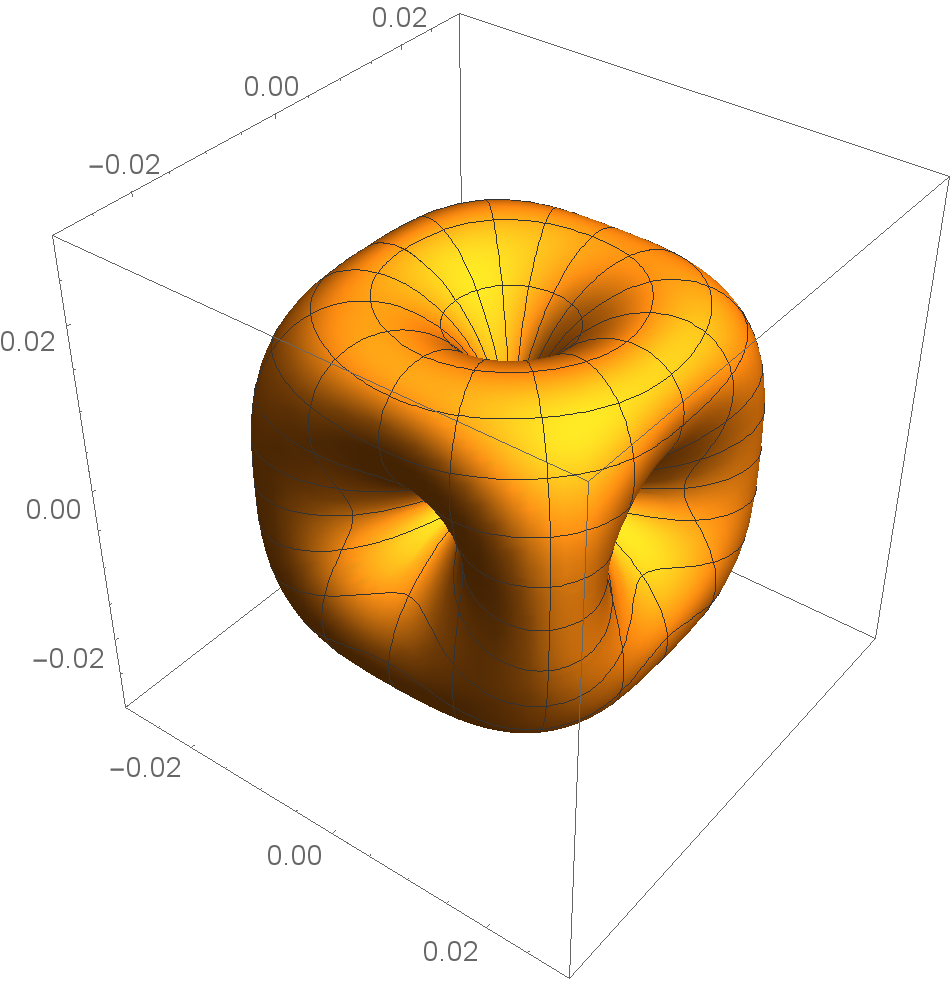}
\caption{A graphical representation of the energy density in the scalar field $\mu \rho^{\varphi}$ in $(x,y,z)$ space for the eight-mass configuration. The value of $\mu \rho^{\varphi}$ is proportional to the distance of the surface from the origin, and we again display the surface at $r=\pi/2$. We have chosen $m=1$ for each
black hole and $\kappa=0.54$ and $\rho^{\gamma}=0$.}
\centering{}\label{ph8}
\end{center}
\end{figure}

Let us now consider a specific arrangement of masses. Consider eight sources, all with equal mass, and all equidistant from each other on a 3-sphere. The positions of these masses is given in Table \ref{table1}, where ($w$, $x$, $y$, $z$) are Cartesian coordinates on a flat embedding 4-space and ($r$, $\theta$, $\phi$) are hyperspherical polar coordinates on the 3-sphere. The scale factor, $\Omega$, and the energy density of the scalar field, $\rho^{\varphi}$, that result from this configuration are displayed in Figs. \ref{om8} and \ref{ph8}.

\begin{table}[h!]
\begin{center}
\begin{tabular}{|c|l|l|}
\hline
\bf{Point}  & \bf{($w$, $x$, $y$, $z$)} & \bf{($r$,
  $\theta$, $\phi$)}\\ 
\hline
$1$ & $(+1,0,0,0)$  &  $\left(0, \frac{\pi}{2}, \frac{\pi}{2}
\right)$ \\
$2$ & $(-1,0,0,0)$  &  $\left(\pi, \frac{\pi}{2}, \frac{\pi}{2}
\right)$ \\
$3$ & $(0,+1,0,0)$  &  $\left( \frac{\pi}{2}, 0, \frac{\pi}{2}
\right)$ \\
$4$ & $(0,-1,0,0)$  &  $\left( \frac{\pi}{2}, \pi, \frac{\pi}{2}
\right)$ \\
$5$ & $(0,0,+1,0)$  &  $\left( \frac{\pi}{2}, \frac{\pi}{2}, 0
\right)$ \\
$6$ & $(0,0,-1,0)$  &  $\left( \frac{\pi}{2}, \frac{\pi}{2}, \pi
\right)$ \\
$7$ & $(0,0,0,+1)$  &  $\left( \frac{\pi}{2}, \frac{\pi}{2}, \frac{\pi}{2}
\right)$ \\
$8$ & $(0,0,0,-1)$  &  $\left( \frac{\pi}{2}, \frac{\pi}{2},
\frac{3 \pi}{2} \right)$ \\
\hline
\end{tabular}
\end{center}
\caption{The positions of eight regularly arranged points on a 3-sphere in the 4-dimensional Euclidean embedding space with hyperspherical polar coordinates.}
\label{table1}
\end{table}

The scale factor in Fig. \ref{om8} appears as a central bulbous region, with a tube that extends away from the origin at the location of each of the masses. Note that there are $6$ spikes rather than $8$ because only the $r = \pi/2$ slice is shown.  These tubes diverge away to infinity, but have been truncated in producing the diagram. In terms of the intrinsic geometry of the space, the tubes in Fig. \ref{om8} actually approach the geometry of the Schwarzschild solution in the vicinity of the sources. If displayed in an embedding diagram, these regions would each correspond to a different asymptotically flat space and would all be connected to the central bulbous region by a throat. The minimum sphere that can be placed within that throat then corresponds to a marginally trapped surface (both inner trapped and outer trapped, as the geometry is time-symmetric). If only one minimal sphere exists within each of the throats, this must then correspond to the position of the apparent horizon of a black hole.

\begin{figure}[t!]
\begin{center}
\includegraphics[width=13.50cm]{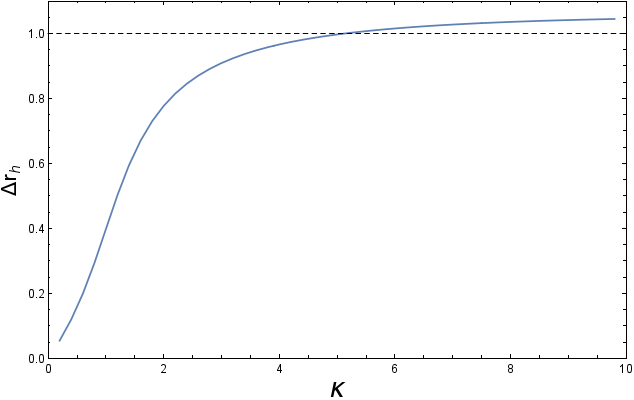}
\caption{The ratio of the apparent horizon size  to the distance to the midpoint between two sources, as determined using the hyperspherical conformal geometry. The position of the apparent horizon is 
shown as a function of $\kappa$. At $\kappa \simeq 5.1$, the apparent horizons from neighbouring black holes touch.}
\centering{}\label{hor8}
\end{center}
\end{figure}

\begin{figure}[h!]
\begin{center}
\includegraphics[width=13.50cm]{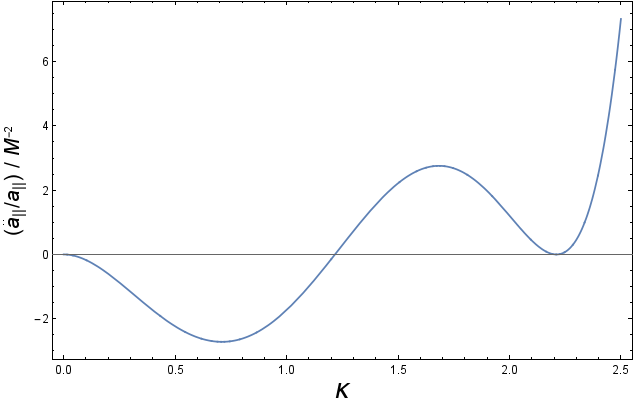}
\caption{The value of $\ddot{a}_{||}/a_{||}$ at the centre of the edge of a lattice cell as a function of $\kappa$, and in units of the total mass of all black holes squared, $M^2$. This quantity becomes positive at $\kappa \simeq 1.2$ and returns to zero at $\kappa \simeq 2.2$.}
\centering{}\label{acc8}
\end{center}
\end{figure}

One of the interesting questions one can ask about this model is whether the horizons of neighbouring black holes touch. This indicates whether the black holes can remain separate as the Universe bounces or are forced to merge. We find the location of the apparent horizon of each of our black holes by calculating the area of a sphere of constant $r$ in a coordinate system centred on the location of the source. This is 
\be
A(r) = \int_0^{2 \pi} \int_0^{\pi} \Omega^4 \sin^2 r \sin \theta \, d\theta \, d\phi \, .
\ee
Once the minimum of this function has been found, then the value of $r$ can be compared to the distance between sources. If the value of $r$ at the apparent horizon exceeds half the distance between sources, then we can say that these black holes have merged. We display the position of the apparent horizon as a function of $\kappa$ in Fig.~\ref{hor8}. The horizons of two neighbouring black holes touch when the two lines cross. This happens at $\kappa \simeq 5.1$. If $\kappa$ exceeds this value, then we expect the black holes to merge before the minimum of expansion.

In order to determine whether these solutions correspond to a bouncing cosmology, we must also determine whether the scale factor of the cosmological region has a positive or negative second time derivative. If this is negative, it corresponds to a maximum of expansion rather than a minimum. Determining the sign of the second derivative of the scale factor is complicated because the space is inhomogeneous. As the cosmological region is close to being (geometrically) spherical, but not exactly so, we must choose a particular curve within the cosmological region and a point along that curve in order to determine the value of the scale factor there.

To determine whether the cosmology is bouncing, we choose a curve that is maximally far from every black hole. If we consider the locus of all points that are closest to any given black hole as the primitive cell of a lattice that covers the conformal 3-sphere, then such a curve corresponds to the edge of one of these cells. The acceleration of the scale factor along this curve is given by
\be
\label{acc1}
\frac{\ddot{a}_{\parallel}}{a_{\parallel}} = \frac{1}{3} \dot{\theta} + \dot{\sigma}(k,k) = - \mathcal{R}(k,k) \, ,
\ee
where $\theta$ and $\sigma_{ab}$ are the expansion and shear of the time-like vector field normal to the initial hypersurface, $\mathcal{R}_{ab}$ is the Ricci tensor of the 3-space, and $k$ is a space-like unit vector tangent to the cell edge. The second equality in this expression has been derived using the evolution equations for $\theta$ and $\sigma_{ab}$, together with the Gauss embedding equation, and the condition $p^{\varphi}=\rho^{\varphi}$.

By direct inspection of the functional form of $\ddot{a}_{\parallel}/a_{\parallel}$ along a cell edge, it can be seen that this quantity takes its minimum value at the midpoint. This is true for all values of $\kappa$. If we require $\ddot{a}_{\parallel}/a_{\parallel}>0$ at all points along a cell edge in order to have a bouncing cosmology, then a necessary and sufficient condition for a bounce is that this condition is true at the middle of the cell edge. To see when this condition is satisfied, we have calculated the value of $\ddot{a}_{\parallel}/a_{\parallel}$ at the midpoint of a cell edge for every value of $\kappa$. The results are displayed in Fig.~\ref{acc8}. It is clear that there is a change in behaviour at $\kappa \simeq 1.2$. For $0<\kappa \lesssim 1.2$, we have $\ddot{a}_{\parallel}/a_{\parallel}<0$. These values of $\kappa$ therefore correspond to solutions in which the scale factor of the cosmological region has a maximum at some point. However, for $\kappa \gtrsim 1.2$, we have $\ddot{a}_{\parallel}/a_{\parallel}>0$ at all points along a cell edge, so it is reasonable to infer that the cosmology bounces if $\kappa \gtrsim 1.2$.

The strange behaviour of the curve in Fig.~\ref{acc8}, in which $\ddot{a}$ returns to zero at $\kappa \simeq 2.2$ after initially increasing, can be understood by considering the sign of the proper mass of each of the black holes. In the region $0 < \kappa  \lesssim 2.2$ the proper mass of each black hole is positive, but it decreases as $\kappa$ increases. At $\kappa \simeq 2.2$, however, the value of $m$ vanishes and it becomes increasingly negative as $\kappa$ increases above $2.2$. When the mass of the black holes vanishes the acceleration of the expansion vanishes, and for negative mass black holes the rapidly increasing acceleration can be attributed to gravity becoming repulsive. If we wish to limit ourselves to the physically relevant case of positive mass black holes, we therefore have an upper bound of $\kappa \lesssim 2.2$. Bounces can occur at higher values of $\kappa$, but they correspond to solutions with negative mass black holes. These features can be understood because $m=0$ when 
\be
3 + \sqrt{1-\kappa} \sin \left[ \sqrt{1-\kappa} \, \frac{\pi}{2} \right] = 0 \, ,
\ee
which is zero when $\kappa \simeq 2.2$. The expression for $\ddot{a}$ contains this factor, as well as a factor $\sin [\sqrt{1-\kappa} \, \pi]$ (which is zero when $\kappa$ is $0$ or $1$) and another more complicated factor which is zero for $\kappa \simeq 1.2$.

By considering both the positions of the horizons, the expansion of the scale factor in the cosmological region, and the condition that the black hole mass must be positive, we therefore have the overall bounds
\be
\label{cond8}
1.2 \lesssim \kappa \lesssim 2.2 \, .
\ee
If $\kappa$ lies above this range, then the black holes have negative mass. If $\kappa$ lies below it, then the magnitude of the scalar field is too small to ensure a bounce and the universe cannot emerge into a new expanding phase. Converting the value of $\kappa$ into the radius of the apparent horizon we find that the bounds in Eq. (\ref{cond8}) correspond to
\be
\label{8bounds}
0.51 \lesssim \Delta r_{\rm h} \lesssim 0.82 \, ,
\ee
where $\Delta r_{\rm h}$ indicates the fractional distance that the horizon extends towards the midpoint between neighbouring black holes, as measured on the conformal hypersphere. The lower bound on $\kappa$ gives the lower bound on $\Delta r_{\rm h}$, and the upper bound on $\kappa$ gives the upper bound on $\Delta r_{\rm h}$. The maximum value of the acceleration occurs approximately midway between the bounds in Eq.~(\ref{cond8}), and takes the numerical value $\ddot{a}_{\parallel}/a_{\parallel} \simeq 2.76 M^{-2}$ (where $M =8m$ is the total mass of all $8$ black holes).

\subsection{All regular arrangements of black holes on a 3-sphere}

In addition to the 8-black hole case considered above, there are five other ways in which it is possible to regularly arrange $N$ sources on a 3-sphere \cite{coxeter}.  These involve $5$, $16$, $24$, $120$ and $600$ black holes, and together with the 8-black hole case, correspond directly to the six regular convex polychora that exist in four dimensions. In this section we carry out a similar analysis to that given above for each of these six possible configurations.

Just as in the 8-mass case, there is a maximum and a minimum allowable value of $\kappa$. The maximum corresponds to the value beyond which the mass of the black holes becomes negative, and the minimum to the value below which there is insufficient energy density in the scalar field to cause a bounce. The latter condition is again determined by insisting that the scale factor at every point along a cell edge is at a minimum of expansion. The logarithm of the numerical values of the maximum and minimum values of $\kappa$ are shown in Table \ref{table2} to three decimal places. Also shown are the minimum and maximum distance of the horizon of one of the black holes, expressed as a fraction of the distance to its nearest neighbours. This corresponds to a generalization of the bounds in Eq.~(\ref{8bounds}), and extends the result to the other five configurations. In the last column of Table \ref{table2} we show the maximum value of the logarithm of ${\ddot{a}_{\parallel} / a_{\parallel}}$.

\begin{table}[t!]
\vspace{30pt}
\begin{center}
\begin{tabular}{|c|c|c|c|c|c|}
\hline
\bf{N}  & ${\bf \log_{10} \kappa_{\rm min}}$ & ${\bf \log_{10} \kappa_{\rm max}}$ & ${\bf \Delta r_{\rm h, min}}$ & ${\bf \Delta r_{\rm h, max}}$ & ${\bf \log_{10} \left[ ( \ddot{a}_{\parallel} / a_{\parallel} ) \, M^2 \right] \vert_{\rm max}}$\\ 
\hline
$5$ & 0.087 & 0.222 & 0.748 & 0.888 & -0.411 \\
$8$ & 0.085 & 0.345 & 0.511 & 0.817 & 0.411 \\
$16$ & 0.201 & 0.542 & 0.453 & 0.888 & 1.100 \\
$24$ & 0.195 & 0.646 & 0.276 & 0.770 & 1.502 \\
$120$ & 0.086 & 1.101 & 0.055 & 0.714 & 3.222 \\
$600$ & 0.503 & 1.590 & 0.066 & 0.910 & 4.401 \\
\hline
\end{tabular}
\end{center}
\caption{Numerical values for the upper and lower bounds on $\kappa$, for the six possible ways of regularly arranging $N$ points on a 3-sphere. Also shown are the corresponding minimum and maximum values of ${\Delta r_{\rm h}}$ and the maximum value of ${\ddot{a}_{\parallel} / a_{\parallel}}$.}
\label{table2}
\vspace{30pt}
\end{table}

\begin{figure}[t!]
\begin{center}
\includegraphics[width=15.0cm]{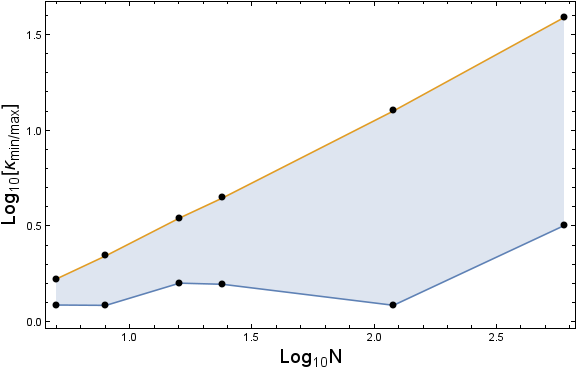}
\caption{The minimum (blue) and maximum (orange) allowed values of $\kappa$ for $5$, $8$, $16$, $24$, $120$, and $600$ regularly arranged black holes on a 3-sphere. The grey shaded region shows allowable values of $\kappa$, where black holes have positive mass and the cosmology still bounces.}
\centering{}\label{results1}
\end{center}
\vspace{30pt}
\end{figure}

\begin{figure}[tbh!]
\begin{center}
\includegraphics[width=13.50cm]{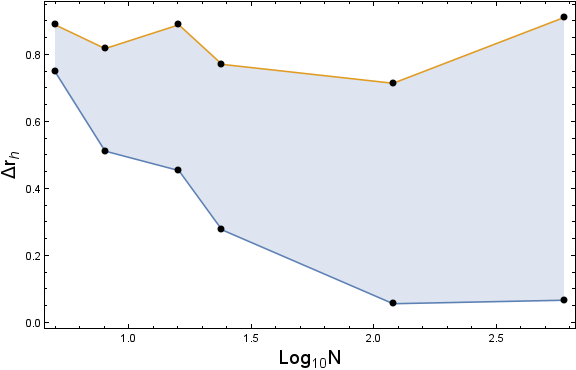}
\caption{The minimum (blue) and maximum (orange) allowed values of $\Delta r_{\rm h}$ for $5$, $8$, $16$, $24$, $120$ and $600$ regularly arranged black holes on a 3-sphere. The grey shaded area shows the allowed region.}
\centering{}\label{results2}
\end{center}
\end{figure}

\begin{figure}[t!]
\begin{center}
\includegraphics[width=13.50cm]{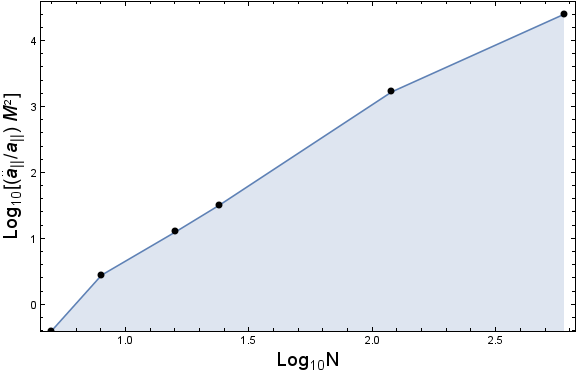}
\caption{The maximum allowed value of $\ddot{a}_{\parallel}/a_{\parallel}$ for $5$, $8$, $16$, $24$, $120$, and $600$ regularly arranged black holes on a 3-sphere. The grey shaded area shows the allowed region, where solutions exist.}
\centering{}\label{results3}
\end{center}
\end{figure}

The results from Table \ref{table2} are shown graphically in Figs. \ref{results1}-\ref{results3}. It can be seen from Fig. \ref{results1} that the upper and lower bounds on $\kappa$ generally increase with the number of black holes, although the lower bound decreases at $N=120$. The grey area in this plot shows the allowed region for $\kappa$, in order that the black hole horizons have positive mass and  the bounce occurs at all points along a cell edge. This band widens as $N$ increases but in a rather irregular way (for the lower bound, at least). This irregularity is probably due to the different configurations having different shaped primitive cells, which is an unavoidable consequence of arranging the sources on the 3-sphere so they are equidistant from all their nearest neighbours.

The fractional distance that the horizon can extend to the midpoint between sources, $\Delta r_{\rm h}$, is shown in Fig. \ref{results2}. The grey region again shows the allowed region for each of the different values of $N$. There is a clear trend, with the minimum allowed value of $\Delta r_{\rm h}$ decreasing as $N$ increases. However, the behaviour of the lower bound in Fig.~\ref{results2} is again a little irregular and it even increases somewhat between $N=120$ and $600$. We again attribute this to the different shaped primitive cells in the different configurations. It is difficult to discern a trend in the upper bound of this plot, which stays around $\Delta r_h \simeq 0.8$ for every configuration. Finally, in Fig. \ref{results3}, we show the allowed values of $\ddot{a}_{\parallel}/a_{\parallel}$. The trend in this case is clear, and increases as the number of masses in the universe is increased.

\section{Black holes in a spatially flat cosmology}
\label{sec:flat}

In vacuum, there do not exist any time-symmetric black-hole lattice solutions with positive mass and zero spatial curvature \cite{num2}. In the presence of a scalar field, however, these solutions do exist. To see this, we can integrate Eq. (\ref{constraint}) over a domain $D$ with periodic conditions and a number of masses $m_i$. In the absense or radiation, this gives
\be
\label{intflat}
\int_D  \frac{\bar{R}}{8} \Omega \sqrt{\bar{g}} \, dV = \frac{\mu}{8} \int_D  \Pi^2 \Omega^5 \sqrt{\bar{g}} \, dV + 2 \pi \sum_{i} m_i \, ,
\ee
where $\bar{g}$ is the determinant of the metric of the conformal 3-space and $dV$ is a volume element on that space. If we now choose the conformal space in Eq. (\ref{le}) to be 3-dimensional Euclidean space, so that
\be
\label{euclid}
d \bar{s}^2 = dx^2 +dy^2 +dz^2 \, ,
\ee
then $\tilde{R}=0$ and Eq. (\ref{intflat}) has solutions if we can ensure
\be
16 \pi \sum_{i} m_i = - \mu \int_D \Pi^2 \Omega^5 \, dx \, dy \, dz \, .
\ee
For $m_i >0$, we require the right-hand side of this equation to be positive, which implies $\Pi \neq 0$ and $\mu <0$. In this section, we therefore assume the conformal Euclidean geometry~(\ref{euclid}) and look for solutions to
\be
\label{master}
\bar{D}^2 \Omega = - \frac{\mu}{8} \Pi^2 \Omega^5 \, ,
\ee
which is what remains of the Hamiltonian constraint (\ref{hc}). We will now study some of the solutions of this equation.

\subsection{Solutions for the conformal factor}

To find solutions to Eq.~(\ref{master}), we choose
\be
\label{piflat}
\Pi^2 = \frac{2 \rho_0}{\Omega^5} \, ,
\ee
where $\rho_0$ is the (constant) energy density in the scalar field at the moment of time symmetry. We are free to make this choice as there is no constraint on the value of $\Pi$ in the initial data (see Section \ref{sec:constraints}). In this case, the Hamiltonian constraint (\ref{master}) becomes linear in $\Omega$, and therefore much easier to solve.

To represent the point-like masses in our lattice, we add the following term to the right-hand side of Eq. (\ref{master}):
\be
m \sum_{n \in \mathbb{Z}^3} \delta^{(3)} ({\mathbf x} - L {\mathbf n}) = \frac{m}{L^3} \sum_{n \in \mathbb{Z}^3} e^{i 2 \pi {\mathbf n} \cdot {\mathbf x}/L} \, .
\ee
This describes a comb of Dirac delta functions separated by distance $L$ and with magnitude $m$. The locations in space where this term is non-zero will eventually correspond to points that will be excised from our geometry, with Eq.~(\ref{master}) being satisfied at all remaining points. We can now look for a solution of the form
\be
\Omega = \sum_{n \in \mathbb{Z}_*^3} A_n e^{i 2 \pi {\mathbf n} \cdot {\mathbf x}/L} + \alpha \, 
\ee
for some constant $\alpha$. Here $\mathbb{Z}_*^3$ is the 3-dimensional set of integers with $(0,0,0)$ removed and the $A_n$ are a set of functions that remain to be determined. We can immediately see that the zero-mode of Eq.~(\ref{master}) gives $m=\frac{1}{4} \mu \rho_0 L^3$, while all other modes give
\be
- \frac{4 \pi^2}{L^2} \vert\mathbf{n}\vert^2 A_n = \frac{1}{4} \mu \rho_0 \, .
\ee
A solution is therefore 
\be
\Omega = - \frac{\mu \rho_0 L^2}{16 \pi^2} \sum_{n \in \mathbb{Z}_*^3} \frac{ e^{i 2 \pi {\mathbf n} \cdot {\mathbf x}/L}}{\vert\mathbf{n}\vert^2} + \alpha \, ,
\ee
which is equivalent to
\be
\label{psisol}
\Omega = - \frac{\mu \rho_0 L^2}{16 \pi^2} \sum_{(n,p,q) \in \mathbb{Z}_*^3} \frac{\cos\left( \frac{2 \pi}{L} n x +\frac{2 \pi}{L} p y+ \frac{2 \pi}{L} q z  \right)}{(n^2+p^2+q^2)} + \alpha \, .
\ee
This solution has singular points at $(x,y,z) = (n L, p L, q L)$ for all $(n,p,q) \in \mathbb{Z}^3$. It also has $\mathbf{v} \cdot \bar{D} \Omega \vert_{\partial D} =0$, where $\mathbf{v}$ is a vector normal to the boundary ${\partial D}$ of a primitive cell of the lattice (again defined as the locus of all points closest to a given mass). These are exactly the properties required for an infinite array of regularly-arranged black holes on a Euclidean 3-space.

\begin{figure}[h!]
\hspace{-40pt}
\includegraphics[width=18cm]{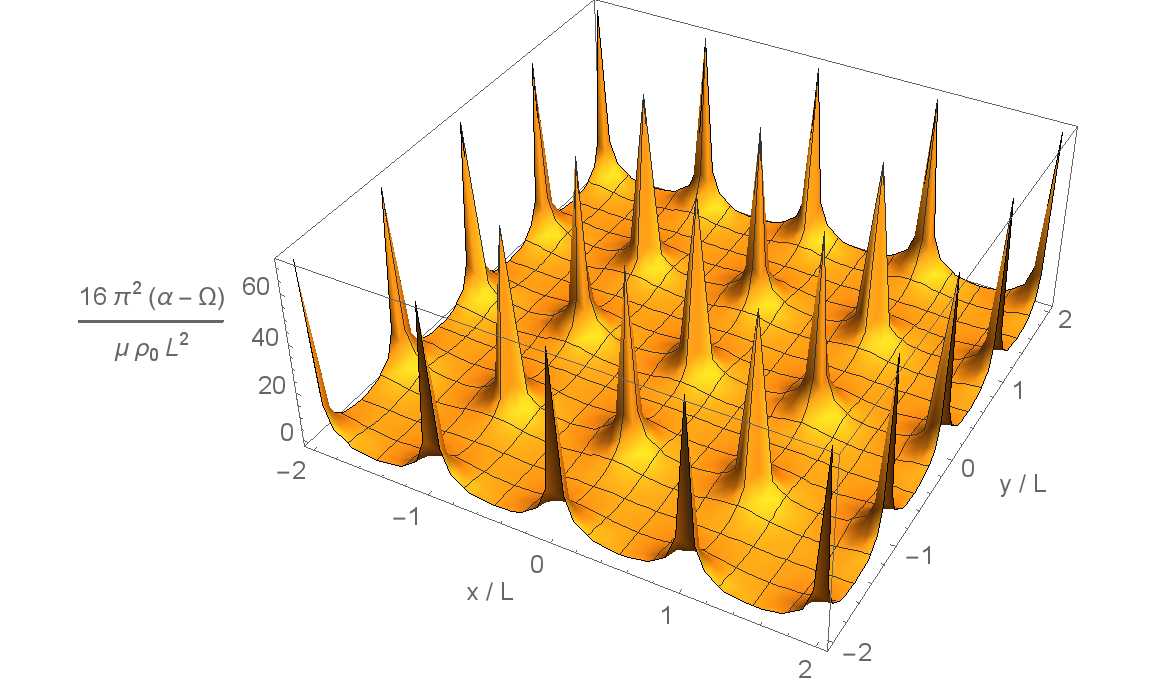} \qquad
\caption{The value of the conformal factor (\ref{psisol}) for a lattice of black holes in a $3$-dimensional Euclidean conformal space. Partial sums from $5 \rightarrow 14$ in $n$, $p$ and $q$ have been averaged over to remove ringing artefacts.}
\centering{}\label{psiplot}
\end{figure}

The functional form of $\Omega$ is illustrated in Fig. \ref{psiplot}. Oscillations arise in taking partial sums in Eq.~(\ref{psisol}), rather than the full (infinite) series. We consider this to be an artefact akin to the Gibbs phenomenon that occurs when modelling a step function using Fourier series. The oscillations have been smoothed out in this figure by taking the average of different partial sums.

\newpage
\noindent
These solutions, corresponding to an infinite number of point sources on a spatially-flat lattice, are characterized by:
\begin{itemize}
\item The distance between sources in the lattice ($L$).
\item The product of the coupling strength and energy density of the scalar field ($\mu \rho_0$).
\item A constant in Eq. (\ref{psisol}) that sets the magnitude of $\Omega$.
\end{itemize}
Once again, it remains to determine the relationship between these quantities and the proper mass of each of the sources.

\subsection{Proper mass of the black holes}

To find the proper mass of the black holes in this space, we will again compare the geometry at a mass point with the corresponding limit of a time-symmetric slice through the Schwarzschild solution. This latter is given in isotropic spatial coordinates by
\be
\label{psis}
\Omega_{\rm S}
= 1+ \frac{m}{2 r} = 1+ \frac{m}{2 \pi} \int \frac{e^{i 2 \pi \mathbf{k} \cdot \mathbf{x}}}{\vert \mathbf{k} \vert^2} d^3 k \, .
\ee
If we define $\mathbf{k} = \mathbf{n}/L$, then our solution can be written as
\be
\label{psi2}
\Omega = - \frac{\mu \rho_0 L^3}{16 \pi^2} \sum_{n \in \mathbb{Z}^3_*} \frac{e^{i 2 \pi \mathbf{k} \cdot \mathbf{x}}}{\vert \mathbf{k} \vert^2} (\Delta k)^3  + \alpha \, ,
\ee
where $\Delta k=1/L$ is the difference between two successive values of $k$ in the sum. In the vicinity of a mass point, it is the small-scale contributions to the sum in Eq.~(\ref{psi2}) that are most important. These contributions are the terms with $n \rightarrow \infty$, so the sum of terms in Eq.~(\ref{psi2}) approaches the following integral expression in the vicinity of a mass point:
\be
\label{psi3}
\Omega \rightarrow - \frac{\mu \rho_0 L^3}{16 \pi^2} \int \frac{e^{i 2 \pi \mathbf{k} \cdot \mathbf{x}}}{\vert \mathbf{k} \vert^2} d^3 k +  \alpha \, . 
\ee
Comparison of Eqs.~(\ref{psis}) and (\ref{psi3}) then gives $\alpha =1$ and $8 \pi m = - \mu \rho_0 L^3$. If we now choose $\mu = - 8\pi$, we have
\be
m = \rho_0 L^3 \, 
\ee
and the proper mass of each black hole is equal to the energy density of the scalar field times the spatial volume of the cell in the conformal space.

\subsection{Location of horizons}

\begin{figure}[tbh!]
\begin{center}
\includegraphics[width=14.0cm]{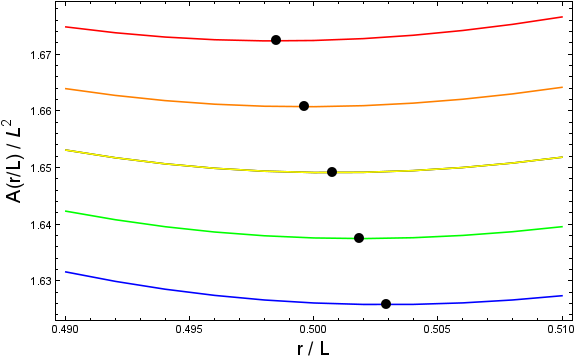}
\caption{The area of spheres of constant $r$ in the conformal Euclidean space. The five curves correspond to $m=0.95 \, L$ (red), $m=0.96 \, L$ (orange), $m=0.97 \, L$ (yellow), $m=0.98 \, L$ (green) and $m=0.99 \, L$ (blue). The minimal value of the area is indicated, in each case, by a black dot.}
\centering{}\label{flatresults1}
\end{center}
\end{figure}

\begin{figure}[t!]
\begin{center}
\includegraphics[width=14.0cm]{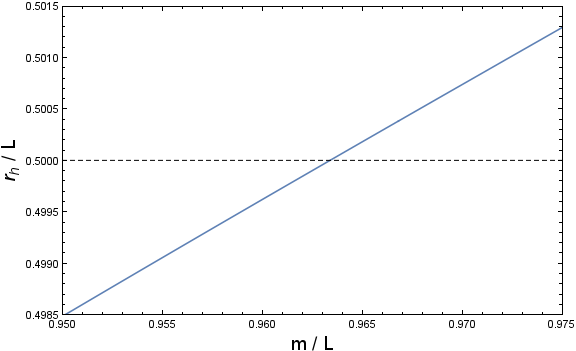}
\caption{The value of the radial coordinate, for a sphere with minimum area, as a function of mass. This gives the location of the apparent horizon, which extends to the midpoint between sources (dotted line) when $m \simeq 0.963 \, L$.}
\centering{}\label{flatresults2}
\end{center}
\end{figure}

Let us now consider the locations of the apparent horizons in these solutions. Again, due to time-symmetry, these are given by minimal closed surfaces in the initial data. We estimate their position by calculating the area of a sphere of constant $r=\sqrt{x^2+y^2+z^2}$ that is centred on a source, and varying the value of $r$ until this area has a local minimum. In this case, the area of such a sphere is
\be
A(r) = \int_0^{2 \pi} \int_0^{\pi} \Omega^4 r^2 \sin \theta \, d\theta \, d\phi \, ,
\ee
where $\Omega$ is given by Eq.~(\ref{psisol}). To find its numerical values, we take partial sums in Eq.~(\ref{psisol}) and put $\alpha = 1$. As discussed above, the partial sum results in Gibbs-like ringing. We therefore take the average of different partial sums to remove this unwanted artefact. The results are displayed in Fig.~\ref{flatresults1} for five different values of $m/L$. Here we take the average of partial sums up to $10$ and $11$ in each of $n$, $p$ and $q$. Also indicated in Fig. \ref{flatresults1} are the positions of the minima. These correspond to the locations of the apparent horizon, whose size increases with $m/L$, as expected. Figure \ref{flatresults2} shows the position of the horizon as a function of $m/L$. It can be seen that the horizons of neighbouring black holes touch when $m$ reaches
\be
m_{\rm max} = 0.963 \, L \, .
\ee
For larger values of $m$, the black holes can no longer be considered as separate objects. This is therefore the maximum mass that can pass through a bounce before the black holes are forced to merge. It is very close to the number that one would obtain from a naive application of the Schwarzschild radius \cite{cc}, which is at $r=m/2$ in isotropic coordinates.

\subsection{Acceleration of cell edges}

Finally, let us determine the acceleration of the scale factor along a cell edge, which is again given by Eq.~(\ref{acc1}). It is straightforward to calculate this quantity for the flat lattice, but problems are introduced by the ringing phenomena discussed above. These are exacerbated in the present case because the acceleration of the scale factor is determined by the Ricci scalar of the space (contracted twice with a space-like unit vector). The derivatives required to calculate the Ricci tensor make the ringing stronger than in  the earlier case.

\begin{figure}[tbh!]
\begin{center}
\includegraphics[width=14.0cm]{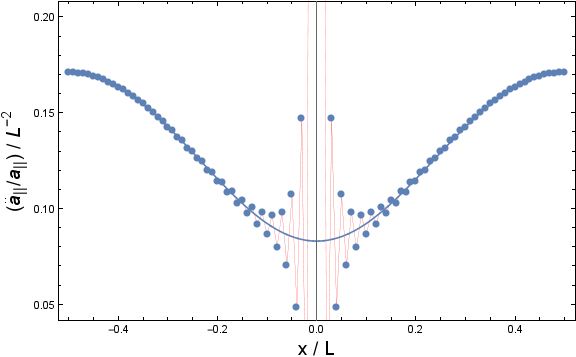}
\caption{The value of $\ddot{a}_{\parallel}/a_{\parallel}$ along a cell edge in the flat lattice. The dots show the value obtained from a partial sum up to $50$ for $n$, $p$ and $q$. The red line that connects them shows the ringing phenomenon from the partial sum. The blue line is a fitted function, as described in the text. We have chosen $m = 0.01 \, L$.}
\centering{}\label{flatresults3}
\end{center}
\end{figure}

One way of reducing the ringing is to increase the number of terms in the partial sum. In this section we therefore increase the maximum value of $n$, $p$ and $q$ to $50$ in Eq. (\ref{psisol}). The results of calculating $\ddot{a}_{\parallel}/a_{\parallel}$ along a cell edge in this case are displayed in Fig. \ref{flatresults3} for $m = 0.01 \, L$. Although the ringing is still present, and gets worse towards the centre of the cell edge, it is reasonably clear what the form of the underlying solution would be if the ringing were absent.

\begin{figure}[t!]
\begin{center}
\includegraphics[width=13.5cm]{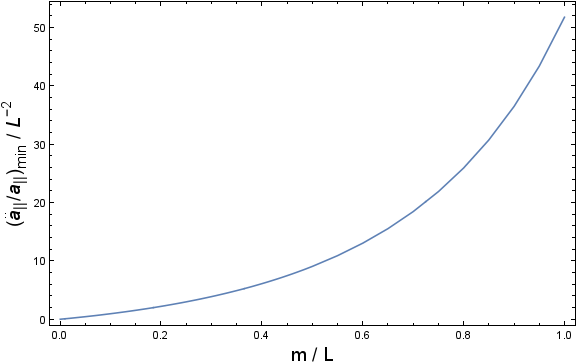}
\caption{The minimum value of $\ddot{a}_{\parallel}/a_{\parallel}$ at the centre of a cell edge as a function of $m$. The function increases monotonically with $m$ and is never negative while the mass is positive.}
\centering{}\label{flatresults4}
\end{center}
\end{figure}

As we want to know the minimum value of $\ddot{a}_{\parallel}/a_{\parallel}$, the ringing presents a particular problem because it is strongest at exactly the centre of a cell edge, where we expect this minumum  to occur. We therefore proceed by fitting sums of sinusoidal functions to the $80\%$ of data points that are furthest from the centre of the cell edge. This allows smooth curves to be fitted to the data and values of $\ddot{a}_{\parallel}/a_{\parallel}$ at the centre of an edge to be determined. Such a fit is shown by the blue curve in Fig.~\ref{flatresults3}, which in this case corresponds to 
\be
f(x) \simeq 0.129 - 0.044 \cos (2 \pi x) - 0.002 \cos^2 (2 \pi x) .
\ee
Repeating this method for different values of $m$ gives Fig. \ref{flatresults4}. This shows the minimum value of $\ddot{a}_{\parallel}/a_{\parallel}$, obtained by fitting smooth curves to the data, and indicates that the acceleration of the scale factor increases monotonically with $m/L$ and without any bound, as far as we can determine. It also appears to be positive for all positive $m$, which means that a bounce always occurs, no matter how low the mass in the black holes gets compared to the mass in the scalar field. We therefore have an upper bound on $m/L$, above which black holes merge, but no lower bound. This is qualitatively different to the hyperspherical cosmologies considered in Section \ref{sec:sphere}.

\section{Black holes in higher-dimensional spaces}
\label{sec:higher}

In some cosmological scenarios one might expect the number of spatial dimensions in the Universe to increase  at sufficiently high densities ({\it {i.e.}} as one approaches the Big Bang). For example, if there were compactified extra dimensions, this would apply at times sufficiently early that the compactification scale exceeds the cosmological particle horizon size.  This would modify both the bounce and merger conditions, so it is interesting to consider the consequences of extra spatial dimensions for the persistence of black holes through a cosmological bounce.

\subsection{Solutions to the constraints}

The constraint equations (\ref{hc}) and (\ref{mc}) are valid for  any number of extra spatial dimensions \cite{witek}. With time-symmetric initial data, and in the absense of radiation, we therefore only require
\be
\mathcal{R}^{(n)} = 2 \mu_n \rho^{\varphi} \, ,
\ee
where $\mu_n$ is a constant and $\mathcal{R}^{(n)}$ is the Ricci scalar of the $n$-dimensional space, the space-time itself being $(n+1)$-dimensional. If we now make a metric ansatz of the form
\be
\label{higherds}
ds^2 = \Omega^\frac{4}{n-2} \left( dr^2 + \sin^2 r \, d \bar{s}^2_{n-1} \right) \, ,
\ee
where $d \bar{s}_{n-1}$ is the line-element of a $(n-1)$-dimensional unit sphere, and we neglect radiation, then the constraint equations are satisfied if
\be
\label{highercon}
\bar{D}^2 \Omega = \frac{1}{4} \left[ n (n-2) - 2  \left( \frac{n-2}{n-1} \right) \mu_n \rho_0 \right] \Omega \equiv \kappa_n \Omega \, ,
\ee
where $\bar{D}$ is now the covariant derivative on the conformal $n$-sphere, $\rho_0 \equiv \rho^{\varphi} \, \Omega^{\frac{4}{n-2}}$ and $\kappa_n$ is defined by the last equality. This reduces to Eq.~(\ref{ham}) if $n=3$ and $\rho^{\gamma}=0$.

If we distribute our scalar field so that $\rho_0$ is constant, then $\kappa_n$ is also constant. The constraint (\ref{highercon}) can then be written in the Helmholtz form:
\be
\Omega_{,rr} + (n-1) \cot r \, \Omega_{,r} = \kappa_n \Omega \, .
\ee
This equation is linear in $\Omega$ for any value of $n$ and has solutions
\be
\label{legendre}
\Omega_i = \alpha_i  \, \frac{\sin^{-\beta}(r) \, P^{\beta}_{\delta} (-\cos r)}{\cos ( \frac{\pi}{2} (1+ 2 \delta )  )} 
+\gamma_i \, \sin^{-\beta} (r) \, Q^{\beta}_{\delta} (-\cos r) \, ,
\ee
where $\alpha_i$ and $\gamma_i$ are constants of integration and 
\be
\beta \equiv \frac{n}{2}-1 \, ,\qquad 
\delta \equiv \frac{1}{2} \left( \sqrt{1+2 \left( \frac{n-2}{n-1} \right) \mu_n \rho_0} -1 \right) \, .
\ee
The functions $P^m_l(x)$ and $Q^m_l(x)$ in Eq. (\ref{legendre}) are the associated Legendre polynomials, and we have included the additional (constant) denominator in the term involving $P^m_l(x)$ so that  we recover a non-zero result in the limit $\rho_0 \rightarrow 0$.

Due to the linearity of Eq. (\ref{highercon}), one can sum any number of solutions of the form (\ref{legendre}) to obtain a new solution. If either or both of $\alpha_i$ and $\gamma_i$ are non-zero, then the solution (\ref{legendre}) diverges as $r \rightarrow 0$. We can therefore take this to be the location of one of the sources in our higher-dimensional lattice. To ensure that the geometry is smooth away from this point, we require the spatial derivatives of $\Omega_i$ to vanish at $r=\pi$. If $n$ is even, this requires $\gamma_i=0$. If $n$ is odd, it requires $\alpha_i=0$. By rotating these solutions on the conformal $n$-sphere and setting the appropriate constants to zero, we therefore have initial data for $N$ arbitrarily located black holes in a positively-curved $(n+1)$-dimensional cosmology.

\subsection{Proper mass of sources}

To find the mass of each of the black holes in these solutions, we follow a similar approach to that in Section \ref{spheremass}. The first step is to transform coordinates so that the conformally spherical geometry in Eq. (\ref{higherds}) becomes conformally flat. Such a transformation is given by $\hat{r} = a \tan (\frac{r}{2})$, so the geometry of our initial surface becomes
\be
\label{higherflat}
ds^2 = \left(  \frac{ (a^2+\hat{r}^2)^{1- \frac{n}{2}}}{2^{1- \frac{n}{2}} a^{1- \frac{n}{2}}}    \Omega \right)^{\frac{4}{n-2}} (d\hat{r}^2 + \hat{r}^2 d \bar{s}^2_{n-1}) \, ,
\ee
where $a$ is a constant. A power series expansion of the overall conformal factor in this equation, around a point-mass at $\hat{r}=0$, then gives
\be
ds^2 =  \left( \alpha + \frac{\beta}{\hat{r}^{n-2}} + \frac{\gamma}{\hat{r}^{n-4}} + \dots  \right)^{\frac{4}{n-2}} (d\hat{r}^2 + \hat{r}^2 d \bar{s}^2_{n-1}) \, ,
\ee
where $\alpha$, $\beta$ and $\gamma$ are constants, and the dots denote smaller terms. Finally, we perform an inversion, to exchange $\hat{r}=0$ with $\hat{r} \rightarrow  \infty$. This can be achieved by the further transformation $r^{\prime} = \delta^2 /\hat{r}$, where $\delta$ is a constant, to  give
\be
\label{hmass}
ds^2 =  \left(\frac{\beta}{\delta^{n-2}} + \frac{\alpha \delta^{n-2}}{r^{\prime \, n-2}} + \frac{\gamma}{\delta^{n-6} r^{\prime \, 2}} + \dots  \right)^{\frac{4}{n-2}} (dr^{\prime \, 2} + r^{\prime \, 2} d \bar{s}^2_{n-1}) \, .
\ee
This can be compared to a slice though the Schwarzschild-Tangherlini solution \cite{Tang}, which is the generalization of the Schwarzschild metric to higher dimensions:
\be
\label{st}
ds^2 = \left( 1 + \frac{2 \pi^{1-\frac{n}{2}} \, \Gamma (\frac{n}{2}) \, m}{(n-1) \, r^{n-2}} \right)^{\frac{4}{n-2}} ( dr^2 + r^2 d\bar{s}^2_{n-1}) \, ,
\ee
where we have used isotropic coordinates \cite{dennison}. Direct comparison of the first and second terms in the conformal factors of Eqs~(\ref{hmass}) and (\ref{st}) then gives the mass of the black hole at $r=0$ as
\be
\label{hmass2}
m = \frac{\alpha \beta (n-1)}{2 \pi^{1- \frac{n}{2}} \, \Gamma (\frac{n}{2})} \, .
\ee
Some comments are required to explain the meaning of this result. If $n>4$, then the two terms that have counterparts in the Schwarzschild-Tangherlini solution are not the largest terms that appear in the conformal factor of Eq.~(\ref{hmass}) in the limit $r^{\prime} \rightarrow \infty$. It can, however, be verified that these larger terms all vanish when $\rho^{\varphi}=0$. This suggests that in higher-dimensional spaces, when the black hole is viewed from infinity on the far side of the Einstein-Rosen bridge,  the leading-order part of the gravitational field comes not from the black hole but from the scalar field, i.e. the energy  in the scalar field appears to gravitate more strongly than the mass of the black hole at large distances.

This situation is rather odd, and does not occur in $(3+1)$-dimensional Einstein-scalar field systems. Nevertheless, it appears to be unavoidable for higher-dimensional spaces and this makes it difficult to distinguish between the gravitational field of the black hole mass and the scalar field. Operationally, we usually define the black hole mass through the leading-order part of its gravitational field at infinity, so if the leading-order part is itself due to the scalar field,  this changes the meaning of the whole procedure. The best we can do is therefore to read off the coefficients in the expansion that look like the terms in the Schwarzschild-Tangherlini geometry and this leads to Eq. (\ref{hmass2}).

One final point is that when the dimension of space is  even, there also appear logarithmic terms in the power series expansion of the conformal factor. These also contribute to the gravitational field at large distances, but do not qualitatively change the above results, as there are no such terms in the Schwarzschild-Tangherlini geometry. In what follows, we use the expression  (\ref{hmass2}) as the proper mass of each black hole in $n$ spatial dimensions. The values of $\alpha$ and $\beta$ in each case will be calculated using the power series expansion of Eqs.~(\ref{legendre}) and (\ref{higherflat}).

\subsection{Acceleration of cell edges in $2\, (n+1)$-mass models}

We now investigate some of the properties of these higher-dimensional spaces, beginning with the acceleration of a cell edge at its midpoint (the cell, as before, being defined as the locus of points closest to each black hole). The acceleration at this point is given by
\be
\label{acc10}
\frac{\ddot{a}_{\parallel}}{a_{\parallel}} = - \mathcal{R}^{(n)}(k,k) \, ,
\ee
which is the same as Eq.~(\ref{acc1}) in the $(3+1)$-dimensional case but with the Ricci tensor now being calculated using the geometry of the higher-dimensional space. This equation can be derived using the higher-dimensional evolution equations \cite{witek}.

\begin{figure}[tbh!]
\begin{center}
\includegraphics[width=15.5cm]{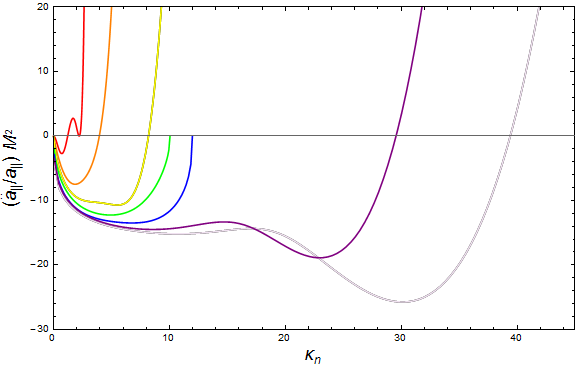}
\caption{The acceleration at the midpoint along a cell edge, for hyper-cubic lattices in spatial dimension $n=3$ (red), $4$ (orange), $5$ (yellow), $6$ (green), $7$ (blue), $8$ (purple), and $9$ (mauve). The red curve is the same as in Fig. \ref{acc8}, and the green and blue curves no longer have real values after $\kappa_6\simeq 10.05$ and $\kappa_7 \simeq 11.98$, respectively. The total mass in each configuration is $M$.}
\centering{}\label{haccplot}
\end{center}
\end{figure}

In order to evaluate this quantity for specific configurations, we consider the tilings on the $n$-sphere corresponding to the $n$-cube. These structures are obtained by placing the largest possible $n$-sphere inside an $n$-cube, so that the sphere touches the cube only at the centre of each of its square faces. If we mark these positions on the $n$-sphere, and then place a black hole at each of them, then each black hole is equidistant from each of its nearest neighbours. The $3$-dimensional example of such a lattice is given by the $8$-mass configuration studied in Section \ref{sec:8mass}. More generally, this procedure leads to a $2 (n+1)$-mass configuration on an $n$-dimensional hypersphere. We also choose the total mass of all of the black holes in each of these spaces to be equal to $M$, so that that they can be easily compared with each other.

The results of calculating the acceleration from Eq.~(\ref{acc10}) for each of the configurations are displayed in Fig.~\ref{haccplot}. The $3$-dimensional results from Fig.~\ref{acc8} are shown in this plot as a red curve. The other six curves correspond to $n=4$, $5$, $6$, $7$, $8$ and $9$. The value of $\kappa_n$ required to make the acceleration positive, and hence cause a bounce, seems to increase with the number of dimensions. For $n=4$, $5$, $8$ and $9$, the curves continue to increase monotonically with $\kappa_n$ after crossing the line that corresponds to $\ddot{a}_{\parallel}/{a}_{\parallel}=0$. Unlike the $n=3$ case, the mass of each black hole ($m$) remains positive for all values of $\kappa_n$. In these dimensions, we therefore only have a lower bound on the value of $\kappa_n$ in order for a bounce to occur.

The curves that correspond to dimensions $n=6$ and $7$ are qualitatively different. They start out with negative acceleration at low $\kappa_n$, just like the other cases, but the value of $m$ becomes zero at the point where $\ddot{a}_{\parallel}/{a}_{\parallel}=0$. For larger values of $\kappa_n$, $m$ becomes negative above $\kappa_6 \simeq 10.05$ and $\kappa_7 \simeq 11.98$, respectively. For values of $\kappa_n$ larger than these bounds, we find that the value of $\ddot{a}_{\parallel}/{a}_{\parallel}$ is no longer real, so it seems that there are no bouncing cosmologies containing black holes of the type given in Eq. (\ref{higherds}) if there are  $6$ or $7$ spatial dimensions. 

\subsection{Locations of horizons in $2\, (n+1)$-mass models}

Let us now consider the locations of the apparent horizons in the same hyper-cubic lattices, in dimensions $n=3$, $4$, $5$, $8$ and $9$. The apparent horizon is, in each case, the smallest sphere that encloses a mass point. As before, we will estimate the position of these surfaces by positioning a mass at $r=0$, and then considering the area of spheres of constant $r$. This area is given by
\be
\label{hah}
A_n(r) = \oint \Omega^{2 \frac{(n-1)}{(n-2)}} \sin^{n-1} (r) \sqrt{g^{(n-1)}} d\theta_1 \dots d\theta_{n-1} \, ,
\ee
where $g^{(n-1)}$ indicates the determinant of the metric of a unit $(n-1)$-sphere, and where the $\theta_i$ are the $(n-1)$-spherical polar coordinates on that sphere. In any given dimension, we look for the value of $r$ that minimizes this quantity.

\begin{figure}[tbh!]
\begin{center}
\includegraphics[width=15.5cm]{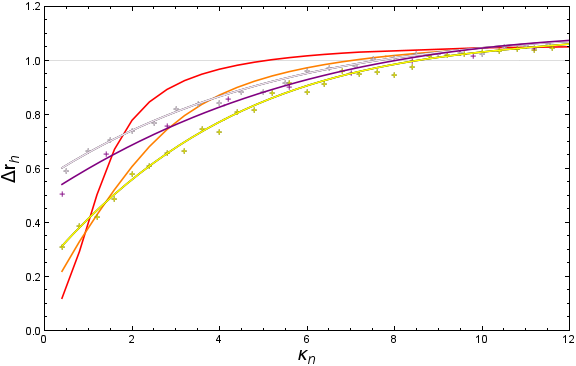}
\caption{The fractional distance from the apparent horizon to the midpoint between two sources, as determined using the conformal hyperspherical geometry of hyper-cubic lattices. Curves show cases with spatial dimension $n=3$ (red), $4$ (orange), $5$ (yellow), $8$ (purple), and $9$ (mauve). The red curve is the same as in Fig. \ref{hor8}, and the crosses show the results from Monte Carlo integrals in the cases $n=5$, $8$ and $9$. The solid yellow, purple and mauve lines are best fitting polynomials of order four, using the method of least squares. The total mass in each configuration is $M$. It is interesting to note that all the curves cross at $\kappa_n \approx 10$.}
\centering{}\label{hhorplot}
\end{center}
\end{figure}

The numerical results we find are shown graphically in Fig. \ref{hhorplot}. As before, we show these results as a fraction of the distance to the midpoint between masses, $\Delta r_h$. When a given curve crosses the line $\Delta r_h =1$ we therefore say that the apparent horizons of neighbouring black holes should be expected to merge. For the lower-dimensional cases, with $n=3$ and $n=4$, it is possible to compute the integrals in Eq. (\ref{hah}) using standard deterministic techniques. However, the computational complexity of such an approach scales very disfavourably as the number of dimensions increases. For dimensions $n>4$ we therefore resort to Monte Carlo integration. This method places points randomly, and computes a statistical approximation to the true value of an integral. It also scales much more favourably with dimensionality, making the necessary numerical integrals possible within a feasible time-scale. The results of the Monte Carlo integrations are shown as crosses in Fig. \ref{hhorplot}. We have fitted curves to this data by performing a least squares fit to polynomial functions of order four, which are also displayed in the figure. 

We find, in each of the cases displayed, that the apparent horizons of neighbouring black holes should be expected to merge when the value of $\kappa_n$ exceeds a certain bound. This bound is lowest for the $3$-dimensional case, for which we obtain $\kappa_3 \lesssim 5.1$. The bound increases by a small amount when an extra dimension is included, but subsequent extra dimensions do not appear to raise it much further (to within the accuracy we can sensibly claim, given our statistical techniques). However, it appears that in every case we have $\kappa_n < 10$. This is not a problem for the cases $n=4$, $5$ and $6$, as there are values of $\kappa_n<10$ in each of these cases that can still produce a bounce (see Fig. \ref{haccplot}). However, this is not true for the cases with $n=8$ or $9$. In both of these cases the lowest value of $\kappa_n$ required to produce a bounce is greater than the bound below which apparent horizons merge. It therefore appears that we cannot produce a bounce in either of these dimensionalities without forcing neighbouring black holes to merge.

Taking both the requirement for a bounce and the requirement for neighbouring black holes to remain distinct consequently appears to place strong restrictions on these solutions in higher-dimensional spaces. In fact, it does not appear that we can produce a bounce if $n>5$ (at least up until $n=9$, which is as far as we have studied here). Inclusion of one or two extra spatial dimensions does, however, remove the upper bound on the number density of black holes that exists in $n=3$, and which comes from the requirement that the mass of each black hole is positive. There is no such bound in dimensions $n=4$ or $5$, and so the black holes can safely be packed in tighter, all the way up until the point their horizons merge.  

\section{Discussion}

In this paper we have considered bouncing cosmological models at the moment of maximum compression. Our models contain a lattice of black holes in a Universe whose energy density is dominated by a scalar field. We then obtained exact solutions for time-evolving models in which multiple distinct black holes persist through the bounce.

First we considered hyperspherical cosmological models.  We found that the model parameter $\kappa$ must lie within a given range of values if the black holes are to have positive mass and the magnitude of the scalar field is to be large enough to ensure a bounce.  A total of six configurations for multiple, regularly-arranged black holes were considered, and it was found that the upper and lower bounds on $\kappa$ generally increase with the number of black holes.  Black holes in a spatially flat cosmology and in the presence of a scalar field were also considered, and were found to allow time-symmetric solutions with positive mass.  We again found {an} upper bound above which black holes merge, but unlike the hyperspherical cosmologies we found no lower bound.

Finally, we considered the bounce and merger conditions on persistent black holes through a cosmological bounce in higher-dimensional spaces, which may be of relevance in cosmological scenarios in which the number of spatial dimensions in the Universe increase at sufficiently high densities. The lower bound on the values of the model parameter $\kappa_n$ required for a bounce seems to increase with the number of dimensions. However, it appears that that there are no bouncing cosmologies containing non-merging black holes of the type discussed in this paper if there are more than five spatial dimensions (i.e. if $n>5$).  We also found that the upper bound on the number density of black holes that exists in $n=3$, {coming} from the requirement that the mass of each black hole {be} positive, is removed when $n=4$ or $5$.

There are several areas in which the current analysis could be extended, and {these}  will be the subject of future work. This includes the study of models in which the black holes persist {without the bounce being time-symmetic, models in which the} black holes merge on the approach to the bounce, and more {detailed} higher-dimensional models. For example, time-asymmetric solutions might be more realistic if the entropy increases at a bounce. Beyond these more mathematical questions, there are also a number of potentially very interesting physical consequences of the models presented here, {and these} are discussed elsewhere \cite{future}.  For example, if there is more than one bounce then it is possible that there are more black holes in each subsequent cycle (i.e. both the newly created ones and the persisting {ones from previous cycles). In this case}, the filling factor of black holes increases with each cycle, so that the bounces may at some point terminate. {More speculatively}, it is possible that BCBHs and PCBHs might act as the seeds for galaxy formation \cite{carr-rees} or even as the dark matter itself \cite{cks}. This offers an interesting alternative to the currently popular suggestion that {PBHs, formed at around} a second after the big bang, might be responsible for these phenomena.

\section*{Appendix A: Multi-Janis-Newman-Winicour Solutions}

Let us now consider solutions with ${\dot{\varphi}=0}$ and $V(\varphi)=0$, which contain a lattice of black holes. These solutions {are perhaps of} less cosmological interest than the solutions presented in Secs.~3,~4, and 5, but {they} constitute interesting generalisations of the scalar field solutions that are often referred to as Janis-Newman-Winicour black holes \cite{JNW4} (although they were discovered by several previous authors \cite{JNW1,JNW2,JNW3}).

If ${\dot{\varphi}=0}$, the momentum constraint (\ref{mc}) is automatically satisfied on time-symmetric hypersurfaces. The Hamiltonian constraint (\ref{hc}) can then be written, in the absence of radiation, as
\be
\mathcal{R} = \mu D^a \varphi D_a \varphi \, ,
\ee
which, for the hyperspherical geometry (\ref{le}), gives
\be
\bar{D}^2 \Omega = \left[ \frac{3}{4} - \frac{\mu}{8} \bar{D}^a \varphi \bar{D}_a \varphi \right] \Omega \,  ,
\ee
where we have taken the conformal metric to be a hypersphere with $\bar{\mathcal{R}}=6$. If we now transform variables from $\Omega$ and $\varphi$ to $\chi$ and $\Psi$, using
\be
\Omega = \vert \chi \vert ^{\frac{1-\beta}{2}} \vert  \Psi \vert ^{\frac{1+\beta}{2}} \, , \qquad 
\varphi = 2 \lambda \ln \left\vert  \frac{\chi}{\Psi} \right\vert  \, ,
\ee
where $\lambda$ is a constant and $\beta = \pm \sqrt{1 - 2 \mu \lambda^2}$, then all constraints are satisfied if
\be
\label{masterX}
{\bar{\nabla}^2 \chi}= \frac{3}{4}  \chi \, , \qquad
{\bar{\nabla}^2 \Psi} = \frac{3}{4} \Psi\, .
\ee
The solutions to these equations are
\be
\chi = \sum_{i=1}^N \frac{\alpha_i}{2 \sin  \frac{r_i}{2} } \, ,  \qquad 
 \Psi = \sum_{i=1}^N \frac{\gamma_i}{2 \sin  \frac{r_i}{2} } \, ,
\ee
where $r_i$ is the $r$ coordinate for the $i$th source in a coordinate system rotated so that the source is at $r=0$. The quantities $\alpha_i$ and $\gamma_i$ are constants to be specified for each of the $N$ terms in the sum.

To determine the proper mass of the black holes in this solution, we consider the limit $r_i \rightarrow 0$. The line-element of the space then becomes
\be 
\label{adm1}
ds^2 \rightarrow \left\vert \frac{{\alpha_i}}{r_i} +A_i \right\vert^{2 (1- \beta)}  \left\vert \frac{{\gamma_i}}{r_i} +B_i \right\vert^{2 (1+ \beta)} d\bar{s}^2 \, ,
\ee
where
\be
A_i  = \sum_{j \neq i} \frac{{\alpha_j}}{2 \sin  \frac{r_{ij}}{2}} \, , \qquad 
B_i  = \sum_{j \neq i} \frac{{\gamma_j}}{2 \sin \frac{r_{ij}}{2} } \, ,
\ee
and $r_{ij}$ is the coordinate distance to the $j$th source in the coordinate system centered on the $i$th source. If we now introduce a new radial coordinate,
\be
r_i^{\prime} \equiv \frac{\vert \alpha_i \vert^{1-\beta} \vert \gamma_i \vert^{1+\beta}}{r_i} \, ,
\ee
then the metric (\ref{adm1}) becomes
\be
\hspace{-20pt}
ds^2 \rightarrow \left( 1+ \frac{2 (1-\beta) A_i }{r_i^{\prime}}  \frac{\vert \alpha_i \vert^{1-\beta} \vert \gamma_i \vert^{1+\beta}}{\alpha_i}  + \frac{2 (1+\beta) B_i  }{r_i^{\prime}} \frac{\vert \alpha_i \vert^{1-\beta} \vert \gamma_i \vert^{1+\beta}}{\gamma_i} \right) d\bar{s}^2 \, .
\ee
This can be compared to the Schwarzschild solution (\ref{schwarz}) in the limit $r \rightarrow \infty$ to find the proper mass of the $i$th source as
\be
\label{mscalar}
m_i = (1-\beta) A_i  \frac{\vert \alpha_i \vert^{1-\beta} \vert \gamma_i \vert^{1+\beta}}{\alpha_i} + (1+\beta) B_i   \frac{\vert \alpha_i \vert^{1-\beta} \vert \gamma_i \vert^{1+\beta}}{\gamma_i}  \, .
\ee
The proper mass is again a function of every $\alpha_i$ and $\gamma_i$.

To determine the scalar charge for the black holes, let us define the scalar charge of a source as
\be
\label{scalarcharge}
q_i \equiv \frac{1}{4 \pi} \int \varphi_{,i} n^i dA \, ,
\ee
where the integral is carried out over a 2-sphere in the limit $r_i^{\prime} \rightarrow \infty$ (or $r_i \rightarrow 0$). In this limit, the radial component of the unit vector $n^a$ is $n^{r^{\prime}} \rightarrow {1- {m}/{r^{\prime}}}$, while the area element is $dA \rightarrow \left( 1+ {2 m}/{r^{\prime}} \right) r^{\prime 2} \sin \theta \, d \theta \, d \phi$. Substituting this into Eq~ (\ref{scalarcharge}) gives the scalar charge of the $i$th source as
\be
\label{qscalar}
q_i = - 2 \lambda A_i  \frac{\vert \alpha_i \vert^{1-\beta} \vert \gamma_i \vert^{1+\beta}}{\alpha_i}  + 2 \lambda B_i  \frac{\vert \alpha_i \vert^{1-\beta} \vert \gamma_i \vert^{1+\beta}}{\gamma_i}  \, .
\ee
Again, this is a function of every $\alpha_i$ and $\gamma_i$. For a given set of $N$ sources, and with a given set of $\alpha_i$ and $\gamma_i$, we can now calculate $m_i$ and $q_i$ for each of them. Note that the solutions given in Section \ref{sec:sphere} have no scalar charge, according to the definition (\ref{scalarcharge}).

The solutions described in this appendix are valid for sources positioned at arbitrary locations on a 3-sphere. It is, however, instructive to compare them to the Janis-Newman-Winicour solution for a static black hole in an asymptotically flat space. This is given in isotropic coordinates by \cite{JNW4}
\be
\hspace{-70pt}
ds^2 = - \frac{\left( 1 - \frac{m}{2 r} \right)^{2 \beta} }{\left( 1 + \frac{m}{2 r} \right)^{2 \beta}} dt^2 + \left( 1 - \frac{m}{2 r} \right)^{2 (1-\beta)}  \left( 1 + \frac{m}{2 r} \right)^{2 (1+\beta)} (dr^2 +r^2 (d\theta^2 +\sin^2 \theta d\phi^2)) \, ,
\ee
with the scalar field being 
\be
\varphi = 2 \lambda \ln \frac{(1- \frac{m}{2 r})}{(1+ \frac{m}{2 r})} \, .
\ee
A coordinate transformation $r \rightarrow K \tan (r/2)$ with constant $K$ gives
\begin{equation*}
\hspace{-70pt}
ds^2 = -\frac{( \tan \frac{r}{2} - \frac{m}{2 K})^{\beta}}{( \tan \frac{r}{2} + \frac{m}{2 K})^{\beta}} dt^2 + \frac{K^2}{4 \cos^4 \frac{r}{2}} \left( 1-\frac{m}{2 K \tan \frac{r}{2}} \right)^{2 (1-\beta)}  \left( 1+ \frac{m}{2 K \tan \frac{r}{2}} \right)^{2 (1+\beta)} 
d\bar{s}^2
\, .
\end{equation*}
The conformal factor of the spatial part of the metric is then
\be
\Omega = \frac{\sqrt{K}}{\sqrt{2} \cos \frac{r}{2}} \left( 1-\frac{m}{2 K \tan \frac{r}{2}} \right)^{(1-\beta)/2}  \left( 1+ \frac{m}{2 K \tan \frac{r}{2}} \right)^{(1+\beta)/2} \, .
\ee
If we want the horizon at $r=m/2$ to appear at $r=\pi/2$ in the new coordinates, we should choose $K=m/2$. This gives
\be
\Omega = \left\vert \frac{\sqrt{m}}{2 \cos \frac{r}{2}} - \frac{\sqrt{m}}{2 \sin \frac{r}{2}} \right\vert^{(1-\beta)/2}  \left\vert \frac{\sqrt{m}}{2 \cos \frac{r}{2}} + \frac{\sqrt{m}}{2 \sin \frac{r}{2}} \right\vert^{(1+\beta)/2}  \, ,
\ee
where the solution is valid both inside and outside the horizon. 

\begin{figure}[t!]
\begin{center}
\includegraphics[width=10.5cm]{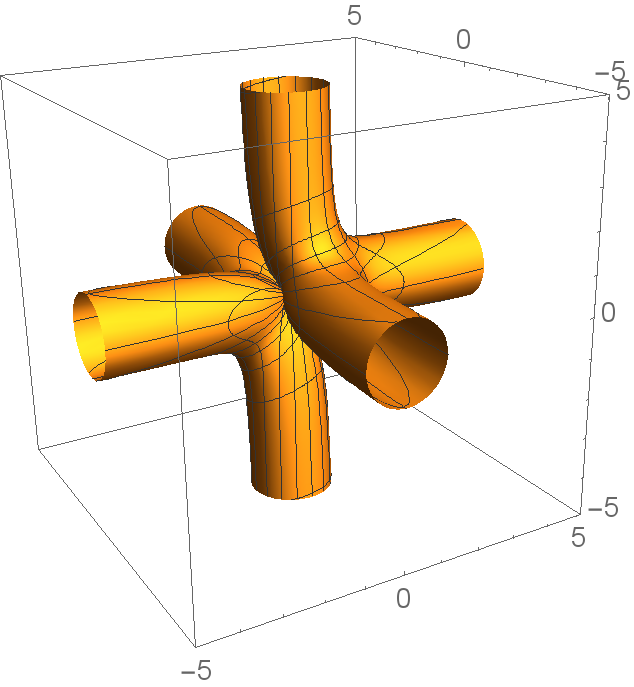}
\caption{The value of the conformal factor, $\Omega$, for the eight regularly arranged Janis-Newman-Winicour-like black holes. The value of $\Omega$ is proportional to the distance of the surface from the origin. We have chosen to taken $m=1$, $\mu=1$, $\lambda=1/2$, and $\beta$ negative to create this plot.}
\centering{}\label{fig2}
\end{center}
\end{figure}

\begin{figure}[t!]
\begin{center}
\includegraphics[width=10.5cm]{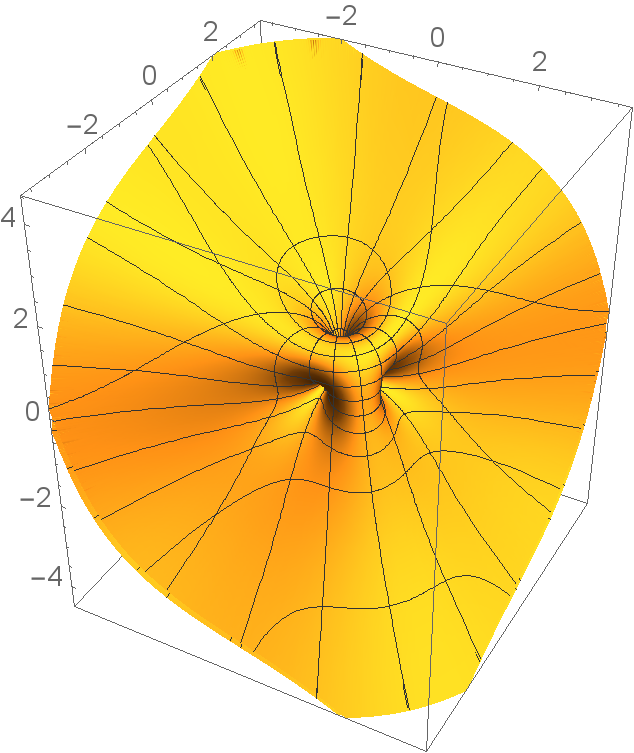}
\caption{The value of the scalar field, $\varphi$, for eight regularly arranged Janis-Newman-Winicour-like black holes. The value of $\varphi$ is proportional to the distance of the surface from the origin. Here we have chosen the same parameter values as in Fig. \ref{fig2}.}
\centering{}\label{fig1}
\end{center}
\end{figure}

It is now clear that the Janis-Newman-Winicour solution corresponds to two black holes on a hypersphere, with $\chi$ and $\Psi$ given by
\be
\chi = \frac{\sqrt{m}}{2 \sin \frac{r}{2}} - \frac{\sqrt{m}}{2 \cos \frac{r}{2}} \, , \qquad
\Psi = \frac{\sqrt{m}}{2 \sin \frac{r}{2}} + \frac{\sqrt{m}}{2 \cos \frac{r}{2}} \, .
\ee
This prompts us to take the following parameters for these two black holes:
\be
\alpha_1 = \sqrt{m} \, ,\quad \alpha_2 = -\sqrt{m}\, , \quad \gamma_1 = \sqrt{m} \, , \quad
 \gamma_2 = \sqrt{m} \, .
\ee
The proper mass and scalar charge on these black holes can then be read off from Eqs. (\ref{mscalar}) and (\ref{qscalar})  as
\be
m_1 =m_2 = \beta \vert m \vert \, , \qquad
q_1=q_2 = 2 \lambda \vert m \vert \, .
\ee
Having considered the 2-mass configuration, which turns out to be isometric to a time-symmetric slice through the Janis-Newman-Winicour solution, let us consider eight regularly arranged masses on a hypersphere (as considered in Section \ref{sec:sphere} for the case $D_a \varphi =0$). For this case we can choose
\begin{eqnarray*}
\alpha_i &=&  +\sqrt{m} , \, \qquad
\gamma_i =  \sqrt{m}  \qquad (i=1,3,5,8) \, , \\  
\alpha_j &=&  - \sqrt{m} \, , \qquad 
\gamma_j =  \sqrt{m} \qquad (j=2,4,6,7)  \, ,
\end{eqnarray*}
which leads to masses and charges 
\be
m_i = [ 3 \sqrt{2} (1+ \beta) + \beta ] \vert m \vert \, , \qquad q_i = 2 (1+3 \sqrt{2}) \lambda \vert m \vert \, .
\ee
Graphical representations of the values of $\Omega$ and $\varphi$ on $r=\pi/2$ slices are given in Figs. \ref{fig2} and  \ref{fig1}, respectively. Using the relevant evolution equations, it is possible to determine that some parts of this space are at a minimum of expansion, while others are at a maximum. It can also be seen that (at least) one circle collapses all the way to zero size. Therefore, although these solutions are interesting generalizations of the initial data of the Janis-Newman-Winicour solution, they do not constitute promising cosmological models.

\section*{Appendix B: Solutions with ${\mathbf V(\varphi) \neq 0}$}

Further solutions with $\dot{\varphi} =0$ can be found by allowing $V(\varphi) \neq 0$. These solutions are of more cosmological interest but present difficulties in terms of their interpretation as a universe filled with black holes. If we take
\be
V (\varphi) = V_0 e^{- \nu \varphi} \, ,
\ee
then Eq. (\ref{conformal}) implies, in the absence of radiation,
\be
\Omega = \vert \chi \vert^{\frac{\nu^2}{(\nu^2 +2 \mu)}} \, ,
\ee
where $\Omega$ is the conformal factor from Eq. (\ref{le}), and $\chi$ obeys
\be
\frac{\bar{\nabla}^2 \chi}{\chi} = \frac{(3 - \mu V_0) (\nu^2+2 \mu)}{4 \nu^2} \equiv \tilde{\kappa}\, .
\ee
If $\tilde{\kappa}<1$, the solution to this equation is
\be
\label{chiV}
\chi = \sum_i \chi_i = \sum_{i=1}^N \left( \alpha_i \frac{\cos (\sqrt{1-\tilde{\kappa}} \, r_i)}{\sin (r_i)} + \gamma_i \frac{\sin (\sqrt{1-\tilde{\kappa}} \, r_i)}{\sin (r_i)} \right) \, .
\ee
If $\tilde{\kappa}>1$, the trigometric functions in the numerators above are replaced by hyperbolic functions. In both cases the solution for the scalar field is 
\be
\varphi = \left( \frac{4 \nu}{\nu^2 +2 \mu} \right)  \ln \vert \chi \vert \, .
\ee
If we now require that each term in the sum be smooth at $r_i=\pi$, then we find
\be
\gamma_i = -\frac{\alpha_i}{\tan (\sqrt{1-\tilde{\kappa}} \, \pi)} \, .
\ee
Eq.~(\ref{chiV}) can now be recast as
\be
\chi = \sum_{i=1}^N \alpha_i \frac{\sin (\sqrt{1-\tilde{\kappa}} \, (\pi-r_i))}{\sin (\sqrt{1-\tilde{\kappa}} \, \pi) \sin (r_i)} \, .
\ee
These solutions are characterized by:
\begin{itemize}
\item Complete freedom in the positions of the sources on a reference 3-sphere.
\item One number to be specified for each black hole ($\alpha_i$).
\item A single number ($\mu$) for the coupling strength of the scalar field.
\item A number ($V_0$) that specifies the magnitude of the potential.
\item A number ($\nu$) that parametrises the potential.
\end{itemize}
While the geometry associated with these black holes is well defined, their proper mass is not. This is because the space on the far side of the Einstein-Rosen bridge of each black hole approaches Minkowski space at a rate $\mathcal{O} \Big( r^{-\frac{\nu^2 + 2 \mu}{\nu^2 - 2 \mu}} \Big)$. There is also a deficit angle in that space, making the process of defining mass highly problematic. We leave the possible further study of these solutions to future work.

\vspace{20pt}
\noindent
{\bf Acknowledgments.}
We thank S.~Beheshti, A.~Graham and R.~Tavakol for helpful discussions and contributions. BC thanks the Research Center for the Early Universe (RESCEU) at the University of Tokyo and the Department of Mathematics and Statistics at Dalhousie University, and AC thanks the School of Physics and Astronomy at Queen Mary University of London, for hospitality received during this work. TC acknowledges support from the STFC and hospitality from {the Department of Mathematics and Statistics at} Dalhousie University. AC acknowledges financial support from NSERC.

\end{document}